\documentclass{ar-1col}
\usepackage{cite}
\usepackage{xspace}
\usepackage{url}
\usepackage{slashed}
\usepackage{soul,upgreek}
\usepackage{amssymb}

\newcommand{\chiDM}{\ensuremath{\chi}\xspace}
\newcommand{\IP}{invisible particle}
\newcommand{\mMed}{\ensuremath{M_{\rm{med}}}\xspace}
\newcommand{\mmed}{\mMed}
\newcommand{\gDM}{\ensuremath{g_{\chiDM}}\xspace}
\newcommand{\gdm}{\gDM}
\newcommand{\gl}{$g_{\ell}$\xspace}
\newcommand{\gdmq}{\ensuremath{g_{\chiDM q}}\xspace}
\newcommand{\gq}{$g_{\mathrm{q}}$\xspace}
\newcommand{\mdm}{\ensuremath{m_{\chiDM}}\xspace}

\newcommand{\ghZprimeZprime}{\ensuremath{g_{Z'Z'h}}\xspace}

\newcommand{\pt}{\ensuremath{p_\mathrm{T}}\xspace}
\newcommand{\pT}{\ensuremath{p_\mathrm{T}}\xspace}
\newcommand{\MET}{\ensuremath{\slashed{E}_T}\xspace}
\newcommand{\Zprime}{\ensuremath{{Z}^\prime}\xspace}
\newcommand{\fb}{\ensuremath{\mathrm{fb}}\xspace}
\newcommand{\pb}{\ensuremath{\mathrm{pb}}\xspace}
\newcommand{\ifb}{\ensuremath{\mathrm{fb}^{-1}}\xspace}

\usepackage[
  bookmarks=false,
  pdfpagelabels=false,
  setpagesize=false,
  implicit=false
]{hyperref}

\jname{Annu. Rev. Nucl. Part. Sci.} \jvol{68} \jyear{2018}
\doi{10.1146/annurev-nucl-101917-021008}

\begin{document}

\markboth{Boveia $\bullet$ Doglioni}{Dark Matter Searches at Colliders}

\title{Dark Matter Searches at Colliders}

\author{Antonio Boveia$^1$ and Caterina Doglioni$^2$
\affil{$^1$Physics Department, The Ohio State University, Columbus, Ohio 43210, USA} \affil{$^2$Fysikum, Division of Particle Physics, Lund
University, 22363 Lund, Sweden}}

\begin{abstract}

Colliders, among the most successful tools of particle physics,
have revealed much about matter. This review describes how
colliders contribute to the search for particle dark matter,
focusing on the highest-energy collider currently in operation,
the Large Hadron Collider (LHC)\ at CERN. In the absence of hints about the character
of interactions between dark matter and standard matter, this review emphasizes what could be observed in the
near future, presents the main experimental challenges, and discusses how
collider searches fit into the broader field of dark matter searches. Finally, it
highlights a few areas to watch for the future LHC program.
\end{abstract}


\maketitle

\tableofcontents

\section{INTRODUCTION}\label{sec:intro}

Cosmological observations of dark matter (DM) are perhaps the most persuasive experimental
evidence for physics beyond the Standard Model (BSM)\ of particle physics.
DM may not be composed of particles at all, but the Standard Model of particle physics
in describing ordinary matter gives us a strong reason to consider a particle description of DM as well. 

Neverthless, the evidence for DM (described in, e.g., Reference~\citen{Bertone:2004pz}), is inconsistent with the properties of any known 
particle. While DM has gravitational interactions with normal matter, DM particles are {dark (see the sidebar titled Particle Properties of\ Dark Matter); nongravitational
interactions of DM must be relatively rare. DM is very stable, with a lifetime comparable to that of the Universe (e.g., \citen{Cohen:2016uyg}).
It is also nonrelativistic and collisionless. 
But most striking is its abundance at the present day [relic abundance~\cite{Ellis:1999mm}],
extracted from measurements of the cosmic microwave background~\cite{Ade:2015xua}.
There is approximately five times as much DM as the matter described by the Standard Model. 
This fact provides one of the few quantitative clues about BSM physics, and suggests that 
the complexity of the DM particle sector could match or exceed that of ordinary matter.

\begin{textbox}
\section{PARTICLE PROPERTIES OF DARK MATTER}

\begin{enumerate} 

\item Darkness: Most DM particle candidates produced in particle collisions are effectively invisible to traditional collider experiments. However, any remaining products of the collision event are not. Invisible particles can be
accompanied by one or more visible recoiling particles, leading to
missing momentum in the transverse plane, whose magnitude is
termed \MET. This is one of the main signatures of DM in
colliders.

\item Very long lifetime: If DM is a particle, it does not seem to decay.
Conservation laws, such as those implied by $Z_2$ symmetry [e.g., \textit{R} parity in supersymmetry (SUSY)], 
can prevent the DM particle from decaying into any
lighter even-parity Standard Model particle. DM particles can also be produced in pairs by the decay of other particles, charged
under the same gauge group as the Standard Model, or singly if the
parent is a color triplet.

\end{enumerate} 

\end{textbox}

Particle physicists are increasingly keen to understand what DM is, if it is indeed composed of particles. 
Some experimenters, using \MakeLowercase{Direct Detection} (DD) experiments, look for Galactic DM
colliding with underground targets made of ordinary matter~\cite{0954-3899-43-1-013001}.
Others, using \MakeLowercase{Indirect Detection} (ID) experiments, search for the products of annihilating DM
concentrated within the gravitational potential wells of the Milky Way and elsewhere~\cite{Gaskins:2016cha}.
If the only interaction between DM and ordinary matter is gravitational, these experiments may never observe
it directly. To succeed, both types of searches require that DM interact
with ordinary matter in some way: DM--nucleon (or DM--electron) interactions
in DD searches or DM annihilation to Standard Model particles in ID searches. 

Colliders, among the most successful tools in particle physics, have revealed much about ordinary matter.  
If DM can be produced at colliders, they will likely remain one of our preferred tools for learning more about it,
regardless of where DM particles are first discovered.
As with DD and ID experiments, collider DM production relies upon the existence of interactions between the
colliding Standard Model particles and the DM particles.
If we can produce DM at the Large Hadron Collider (LHC)~\cite{LHC2008}
or its successors, we might begin to comprehend the forces that connect 
ordinary matter to DM, and to understand how the two interacted
shortly after the Big Bang, leading to the Universe we see today.

This review describes how experiments at particle colliders contribute to the search for DM,
focusing on the ATLAS, CMS, and LHCb experiments~\cite{ATLAS2008,CMS2008,LHCb2008} at the highest-energy
collider currently in operation, the LHC at CERN.
The LHC results presented in this review include up to
36~\ifb of proton--proton collision data recorded through 2017 during 
the 2015--2018 LHC run at 13-TeV center-of-mass energy (Run 2).
This data set is almost twice as large as what was used for the Higgs discovery
at 7- and 8-TeV center-of-mass energy (approximately 20~\ifb) during 2010--2012 (Run 1),
but it comprises only 1\% of the 3,000~\ifb expected with the full High-Luminosity LHC (HL-LHC)
run, planned to start in 2026. 

Given the absence of any hints as to the character of DM--Standard Model
matter interactions, this review emphasizes what could be observed in the near
future, presents the main experimental challenges, and discusses how
collider searches fit into the broader field. Finally, it
highlights a few areas to watch for the future LHC program.

\section{REACTIONS FOR INVISIBLE-PARTICLE SEARCHES AT THE LHC}\label{sec:02_Reactions}

Just as neutrinos do, 
DM produced at colliders would almost always pass invisibly through the detector. 
In this section, we describe DM
production from a pragmatic, collider physicist's perspective,
focusing on a selection of simple models with distinct and
testable LHC signatures.\footnote{For other perspectives, see Literature Cited.} Moreover, we use the term {invisible
particles} (rather than DM) when emphasizing that detecting such
particles need not be a discovery of DM.\footnote{For example,
{\IP}s may decay after leaving the detector, a decay that is
essentially prompt on cosmological timescales.}

The body of DM model literature can be divided into two extremes.
Fully specified, self-consistent models such as SUSY provide
specific features that can be exploited for narrowly targeted
searches, while simplified models with a few components can
capture broad collider signatures of classes of models, serving as
benchmarks for more general but less optimal searches. Key to both
are the determinative details of the interactions between DM
and ordinary matter, rather than DM itself.

To restrict the scope of this review, we emphasize 
(\textit{a}) models where the DM has an effective interaction with Standard Model particles
such that it can be produced in colliders; 
(\textit{b}) models that include $Z_2$ symmetry to stabilize DM; 
(\textit{c}) models connecting to the most commonly studied cosmological history, 
where it is assumed that DM is in thermal equilibrium in the
early Universe and freezes out to the current abundance [see Reference~\citen{Steigman:2012nb} for an overview; we briefly comment on others with alternate cosmological histories, which also
have interesting signatures~\cite{Bernal:2017kxu,Brooijmans:2018xbu,Evans:2017kti}]; 
(\textit{d}) models in which the DM is a Dirac fermion; and
(\textit{e}) models that mimic the pattern of flavor violation found in the Standard Model, 
referred to as minimal flavor violation (MFV)~\cite{DAmbrosio:2002vsn}. 
These are the models used in LHC Run 2 searches. 
Departures from these assumptions are discussed in~Reference~\citen{Abercrombie:2015wmb}.

\subsection{Higgs and {\bf \it{Z}} Boson Portals}\label{sec:HZPortalModels}

\begin{figure}[!htpb]
\includegraphics[width=\textwidth]{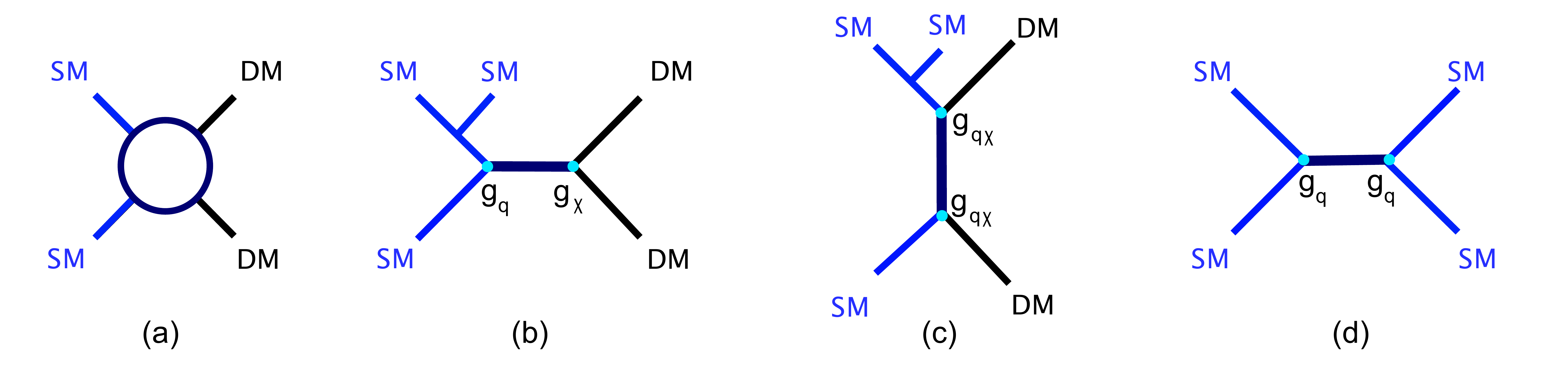}
\caption{(\textit{a}) The interaction between DM\ and Standard Model particles via an unspecified interaction (e.g., an EFT).
(\textit{b}) Examples of simplified model processes where the interaction is mediated by an intermediate particle (with additional radiation off one of the initial-state quarks). 
(\textit{c}) The same model, in which  the mediator decays back into Standard Model particles, with coupling constant  \gq  for the mediator--quark--quark vertex and constant  \gdm for the mediator--DM vertex. 
Abbreviations:\ BSM, beyond the Standard Model; DM, dark matter; EFT, effective field theory; SM, Standard Model. \label{fig:feynman_0}}
\end{figure}

Extending the Standard Model with a single DM particle, one
may arrive at models where the Higgs boson or the \textit{Z} boson mediates
the DM--Standard Model\  interaction, also called portal models. 
The Higgs boson and the \textit{Z} boson in these models the are examples of {mediators}, 
particles governing the DM--Standard Model interaction. 
Figures~\ref{fig:feynman_0}\textit{b,c} shows example Feynman diagrams.

{Higgs portal} models~\cite{Patt:2006fw,Djouadi:2011aa} can be constructed by adding only DM and no other new particles to the Standard Model. 
Only the most recent recent generations of collider and DD
experiments have reached the energies and luminosities necessary to
search for such DM. Direct collider searches for the
invisibly decaying Higgs bosons are augmented by measurements of
other Higgs properties, which can be very sensitive to couplings
to new particles.

{\textit{Z} portal} models are strongly constrained
by LEP and DD experiments (described in Reference \citen{Escudero:2016gzx} and Section~\ref{sec:results_ZHSearches}). 
Either of these mediators (the \textit{Z} or Higgs boson) is light in
comparison to the LHC energy and can be produced on-shell, so
collider searches may still constrain these models 
through precision studies of the visible decays of the \textit{Z} and Higgs bosons, 
even if the {\IP}s are much heavier and invisible decays of the mediator
are absent. 

\subsection{Effective Field Theories and Simplified Models of Beyond-the-Standard-Model Mediators}\label{sec:BSMMediatorModels}

More complex than the \textit{Z} and Higgs portal
models are models in which the mediator of the
DM--Standard Model interaction is also a new particle, such
as a heavier version of the $Z$ boson (a $Z^\prime$) or an
additional scalar.

\subsubsection{Effective field theories}\label{sub:EFT}

In some situations, such as when a BSM mediator is heavy compared
with the collision energy, the DM--Standard Model interaction appears to be a
contact interaction. All observables are completely determined by
one rate parameter, the contact interaction scale, that controls
the production rate, and a Lorentz structure, which has a
modest effect on the transverse momentum (\pt) distributions of the
{\IP}s. In this case, effective field theories
(EFTs)~\cite{Beltran:2010ww,Goodman:2010ku,Bai:2010hh,Fox:2011pm} describe the production of {\IP}s. Figure~\ref{fig:feynman_0}\textit{a} depicts an EFT
process. One may
hope that such a description is sufficient for the LHC; the
unknown high-energy details of a complicated interaction are
conveniently integrated out. Moreover, since EFTs do not fix a
mediation mechanism, they provide a framework to systematically
explore a wide range of possible physics.

If, instead, the interaction physics is kinematically accessible
(e.g., the mediator mass is within reach of the typical momentum
transfer in the collision), one should replace the EFT description
with a model specifying further details of the DM--Standard Model matter
interactions~\cite{Shoemaker:2011vi}. Without those details,
however, one can still use the EFT language to obtain results for
later reinterpretation once its high-energy completion is known~\cite{Racco:2015dxa,Busoni:2013lha}.

\subsubsection{Simplified models}\label{sub:simplifiedModels}

When the collision energy is near or higher than the mediator
mass, complementary avenues to study the mediating interaction
develop, analogous to the transition from the Fermi model of weak
interactions at low energies to the Standard Model at higher
energies. For example, at the LHC, a heavy neutral \Zprime
mediator would often decay into the partons that produced it, and
fully reconstructing such visible decays could provide more
information about the interaction than the invisible decays alone.
One can construct simple descriptions of collider phenomenology without 
including the details of additional physics at energies higher than 
the collider scales, and therefore not relevant at the LHC.
These descriptions are termed simplified models
(e.g.,~\citen{Alwall:2008ag, Alves:2011wf}),
and can be considered an intermediate step the between simplest portal
models and full theories. 

Under the assumption that only a few new particles
will be important in the early phase of discovery, 
simplified models can be developed for tree-level pair production of {\IP}s. The set of such models
currently employed for ATLAS and CMS searches is described in
Reference~\citen{Abercrombie:2015wmb}, which builds upon much research from
the wider DM community
(e.g.,~\citen{Fox:2011pm,DiFranzo:2013vra,Abdallah:2015ter}). Although these simplified
models often have to be embedded in a larger theory to satisfy theory
constraints~\cite{Kahlhoefer:2015bea}, in many cases they are sufficient to
describe the leading-order collider phenomenology.

\begin{marginnote}[]
\entry{The ATLAS/CMS Dark Matter Forum}{provided the reference
implementations of the models in Reference~\citen{Abercrombie:2015wmb}
(see \url{https://github.com/LHC-DMWG}})
and implemented in models for various event generators [e.g.,
DMSimp~\cite{Backovic:2015soa}]
\end{marginnote}

%\begin{marginnote}[]
%\entry{The ATLAS/CMS Dark Matter Forum}{provided the reference
%implementations of the models in Reference~\citen{Abercrombie:2015wmb}
%(see \url{https://github.com/LHC-DMWG/model-repository}})
%and implemented in models for various event generators [e.g.,
%DMSimp~\cite{Backovic:2015soa}]
%\end{marginnote}

The most common models include neutral mediator particles
singly produced at the LHC and decaying both to pairs of {\IP}s and to
pairs of Standard Model particles (Figure~\ref{fig:feynman_0}{\em b},{\em
c}). These two-body mediator decays offer simple, attractive benchmarks.
Colored mediators allow vertices involving only one DM particle
and phenomenology akin to that of SUSY models with a squark
mediator~\cite{Papucci:2014iwa,An:2013xka,Bell:2012rg}.
For additional decay signatures, for so-called dark sectors of many
additional particles, and for LHC data sets far larger than at
present, many more simplified models become interesting.
Models of BSM mediation can be classified according to the spin of
the mediator: spin-1 vector or axial--vector mediators (\Zprime),
scalar mediators (referred to as $\phi$ below), and spin-2
mediators~\cite{Han:2015cty}.

{Massive color-neutral spin-1 bosons with vector or
axial--vector couplings} are nearly ubiquitous in BSM theories, so
\Zprime bosons as the mediators connect with a wide class of
models~\cite{Shoemaker:2011vi}. Since the \Zprime coupling to
quarks must be nonzero for its production at the LHC, both
invisible and dijet signatures are discovery channels. This
coupling (or loop-level coupling) to Standard Model partons is also required
for nuclear recoils in underground DM searches.

The models in use at ATLAS and CMS contain vector,
axial--vector, or mixed couplings to quarks and a single species of
{\IP}. The couplings of the \Zprime boson (\gq to all quarks, \gl to
leptons, and \gDM to {\IP}s), the mass of the invisible particle \mdm, and the
\Zprime mass \mmed are free parameters. Lepton decays, if not
included explicitly at tree level, arise through the quark
coupling at loop level (see Reference~\citen{Albert:2017onk} and
references therein). Decays of the spin-1 mediator into neutrinos
are also required by gauge invariance, and add an invisible decay
channel that can enhance signatures of missing transverse
momentum, depending on the size of the
couplings~\cite{Albert:2017onk}. The spin structure of the \Zprime
couplings does not significantly change the LHC phenomenology, but it
has a much greater effect in signals in noncollider searches.
Figure~\ref{fig:feynman_0}\textit{b,d} depicts an example process for this model for the case of invisible and visible decays, respectively. 
The rate of visible decays will increase quickly with increasing \gq.
In order to show the interplay between the constraints from visible
and invisible searches in different decay channels of the mediator,
LHC searches adopt different benchmark coupling scenarios
(described in Reference~\citen{Albert:2017onk} and discussed in Section~\ref{sub:comparisonVisibleInvisible}),
where the coupling to DM is set to unity, the coupling to quarks is set to either 0.25 or 0.1, 
and the coupling to leptons is set to 0, 0.1, or 0.01. 
Those choices are made on the basis of the current LHC search sensitivity
(see Section~\ref{sub:comparisonVisibleInvisible}). These coupling values also 
ensure that the model is still perturbative in most of the parameter space
tested by LHC searches (for a discussion of unitarity and gauge invariance for these models, see Reference~\citen{Kahlhoefer:2015bea}) and that the mediator width is small compared with its mass. 

Models mediated by a \Zprime boson can include additional couplings of the
\Zprime boson to acquire mass through a new baryonic Higgs boson,
$h_B$~\cite{Carpenter:2013xra}, through a coupling \ghZprimeZprime. 
This model variant collapses to the simpler vector model
above in the limit of very heavy \Zprime boson mass. 
Figure~\ref{fig:feynman_1}\textit{a} shows an example Feynman diagram for this case. 
These models can also be
embedded in a type II two--Higgs doublet model
(2HDM)~\cite{Berlin:2014cfa}, as shown in Figure~\ref{fig:feynman_1}\textit{b}.

With appropriate values of the model parameters, 
these models can satisfy the relic density
constraints~\cite{Chala:2015ama}. However, taken in isolation,
the axial--vector model is nonrenormalizable, and without additional ingredients 
perturbative unitarity is violated in certain regions of the
parameter space~\cite{Chala:2015ama,Kahlhoefer:2015bea,Boveia:2016mrp}.

\begin{figure}[!htpb]
\includegraphics[width=\textwidth]{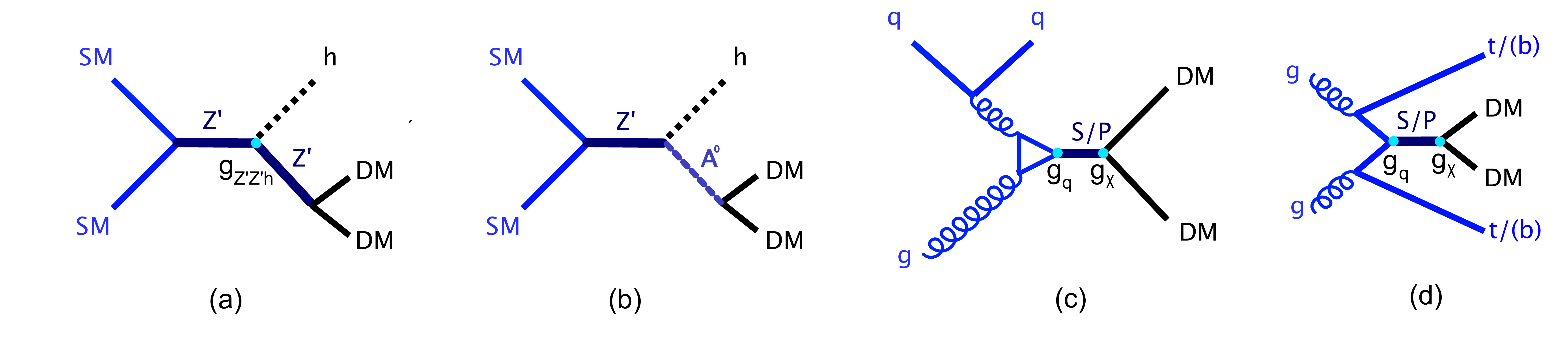}
\caption{
(\textit{a}) Example of a process including baryonic coupling between a vector mediator $Z'$ and an SM Higgs boson. The $Z'$--Higgs coupling is denoted \ghZprimeZprime. 
(\textit{b}) Example of a process from a \textit{U}(1) $Z'$ boson embedded in a 2HDM, where a vector $Z'$ decays to a pseudoscalar $A^0$ that in turn decays to DM particles. 
(\textit{c,d}) Examples of a simplified model process where the interaction is mediated by an intermediate scalar or pseudoscalar particle. In panel \textit{c}, the SM--scalar interaction proceeds through a gluon loop \cite{Haisch:2013ata}, whereas in panel \textit{d}, the pseudo(scalar) is produced in association with a pair of heavy-flavor quarks. The coupling constants that are prefactors to the Yukawa couplings in the model are denoted \gq for the mediator--quark--quark vertex and \gdm for the mediator--DM vertex. 
Abbreviations:\ $A^0/P$, pseudoscalar bosons; $b$, bottom quark; DM, dark matter;  $g$, gluon; $h$, SM Higgs boson; $S$, heavy scalar boson; SM, Standard Model; $t$, top quark; $Z'$, vector mediator; 2HDM, two--Higgs doublet model. }
\label{fig:feynman_1}
\end{figure}

Models including {color-neutral scalar and pseudoscalar mediators (referred to as scalar mediators below)} are analogous
to the Higgs portal model, but with a BSM mediator. Figure~\ref{fig:feynman_1}\textit{c},\textit{d} shows example Feynman diagrams for these cases.  In comparison to the \Zprime
models, a scalar mediator
model~\cite{Buckley:2014fba} has some additional peculiarities.
Under MFV, the couplings of the scalar bosons to fermions
are mass dependent. As with the Higgs boson, there are three
consequences: (\textit{a})\ mediator production through loop-induced
couplings to gluons~\cite{Haisch:2015ioa} and associated with
heavy-flavor quarks~\cite{Buckley:2014fba}, (\textit{b})\ production
cross sections that are smaller than those for vector mediators, and (\textit{c})\ visible
decays that are dominantly to third-generation quarks.
Despite lower production cross sections, the lower backgrounds for these
experimental signatures enable these models to be tested during
LHC Run 2.

The collider phenomenology of the scalar models used by ATLAS and CMS is fully
determined by the masses of the \IP\  and the mediator, the
$\phi$--\IP\  coupling (\gdm), and the $\phi$--fermion (\gq) coupling.
According to the convention used in~Reference~\citen{Abercrombie:2015wmb}, \gq is a
prefactor to the Yukawa couplings to fermions and is set equal for
all quarks. For the same model parameters, the scalar and
pseudoscalar models predict similar kinematic distributions at the LHC.

When introducing an additional scalar, one must consider how this new
scalar relates to the Higgs boson. For example, large mixing with the Higgs can
lead to strong constraints from Higgs measurements, 
when the scalar couples to DM through a Higgs
portal~\cite{Berlin:2014cfa}. If the mediators are pure Standard Model
singlets, then the model is not invariant under
$SU(2)_L$ at the collider scale~\cite{Bell:2016ekl}. Mixing
with the Higgs sector is an example of how gauge invariance can be restored for these models. 
Couplings to the electroweak
gauge bosons can also be added as a consequence of electroweak
symmetry breaking~\cite{Bauer:2016gys,Englert:2016joy}. The
tree-level signatures in this case include Higgs or vector bosons
plus missing transverse momentum and, if the {\IP}s are
sufficiently light, invisible decays of the Higgs boson.

{Colored scalar bosons} allow direct
coupling between Standard Model particles carrying color and {\IP}s carrying $Z_2$
charge~\cite{Bai:2013iqa,Papucci:2014iwa,An:2013xka,
Bell:2012rg,Ko:2016zxg}.
Colored mediators can have a broader set of
multijet signatures and kinematic features than the neutral
mediator models, including the radiation of vector bosons by the
mediator~\cite{Bell:2012rg}.

In colored scalar models, the mediator must be heavier
than the \IP\ so as  to ensure invisible-particle stability. For the current LHC results,
the coupling between {\IP}s and quarks (\gdmq), the invisible-particle  mass, and
the mediator mass are free parameters.

The exchange of a scalar colored under \textit{SU}(3) is analogous to
squarks in the minimal supersymmetric Standard Model\ (MSSM), in which only squarks and neutralinos are light.
In the MSSM, the coupling between DM and the squark is constrained
to be small \cite{Abercrombie:2015wmb}. Without the requirements
of a SUSY framework, this coupling need not be small. For example,
if DM is a standard thermal relic, the couplings required to
obtain the correct DM density are generally higher than
those used by SUSY models. 

\subsubsection{Less-simplified models}\label{sec:LessSimplifiedModels}

Simplified models capture the typical features found in many models. 
As such, they can guide the design of generic searches, but may fail to describe
the full complexity of possible collider signatures that
arise in more complete models. By contrast, 
relying too heavily on a small sample of complete models risks focusing
searches too narrowly on an unrepresentative set of signatures.

To solve this problem, 
many ``less-simplified'' models explore features that arise in special classes of models, 
finding a middle ground between too
simplistic and unnecessarily complex. Because the
set of such models grows quickly with the number of components,
and because there is no broad consensus on which models should be a
priority, very few of them have been explicitly considered by LHC
searches. Here, we highlight a few such models with signatures different from those of
 the simplified models described above.

\begin{figure}[!htpb]
\includegraphics[width=\textwidth]{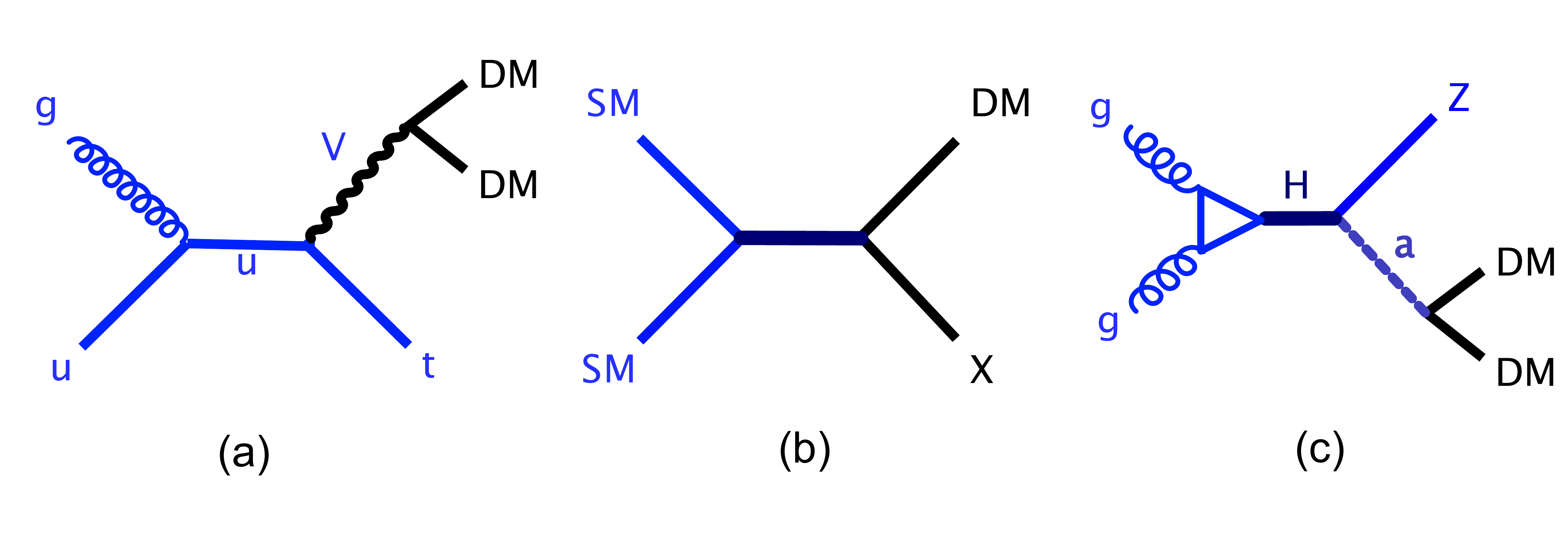}
\caption{
(\textit{a}) Example of a process leading to a single top signature, proceeding through the coupling of a $u$ and a $t$ with a new vector boson, decaying to DM particles. 
(\textit{b}) Example of a collider diagram from a coannihilation model, where two DM particles are present in the final state (one denoted DM and the other  $X$). 
(\textit{c}) Example of a diagram from a 2HDM process, with an interaction between an  $H$, an SM $Z$ boson, and an $a$ mediating the SM--DM interaction. 
Abbreviations:\ $a$, pseudoscalar boson; DM, dark matter; $g$, gluon; $H$, heavy Higgs boson; SM, Standard Model; $t$, top quark;  $u$, up quark; $V$, vector mediator; $X$, coannihilating DM partner, 2HDM, two--Higgs doublet model.}
\label{fig:feynman_2}
\end{figure}

Models with more complex flavor violation structure than the Standard Model 
are just as motivated as those assuming MFV. 
But when constructing viable non-MFV models one must carefully evaluate 
many experimental constraints on flavor-violating processes~\cite{Blanke:2017tnb}. 
Mediators that couple to
DM and a top quark appear in one category of flavor-violating
model that remains least constrained by low-energy
measurements~\cite{Boucheneb:2014wza}. These yield a distinct
``monotop'' LHC signature, as shown in Figure~\ref{fig:feynman_2}\textit{a}.

{Coannihilation} models add two species of dark sector
particles with similar masses (see References~\citen{Buschmann:2016hkc} and \citen{Khoze:2017ixx} for examples). 
The interaction between these two states drives the cosmological
history~\cite{Ellis:1999mm}, as processes involving both types of
particles can efficiently annihilate into Standard Model particles. LHC
signatures include missing transverse momentum accompanied by multiple jets
and/or by the decay products of the 
additional resonant particles in the model in addition to the invisible particle signatures,
as in Figure~\ref{fig:feynman_2}\textit{b}.
The signatures can be very diverse, encompassing some typically
considered in searches with very different motivation than DM (e.g., searches for
leptoquarks). In some cases, these signatures have not received much attention from any
current LHC search~\cite{Buschmann:2016hkc}.

Ultimately, we do not yet know whether the Standard Model Higgs boson is alone 
in the scalar sector, nor whether a single scalar mediator
encodes all of the important features of the complicated
phenomenology of more {complex scalar sectors}. 
A step beyond the simple scalar mediator model, dictated by gauge invariance, 
is to take mixing between this mediator and the Standard Model Higgs boson
into account~\cite{Bauer:2016gys,Berlin:2014cfa}. A much larger
step beyond that is to consider an extended Higgs sector such as a
2HDM, in which one or more of the scalars
acts as the mediator~between DM\ and ordinary matter \cite{Bauer:2017ota,Goncalves:2016iyg,Bell:2016ekl}. 
In such models, the new mediator mixes with the Higgs partners
rather than with the Standard Model Higgs boson, so the model remains
compatible with Higgs measurements. 
Some models developed for LHC searches focus on one Yukawa structure (type II)~\cite{Pich:2009sp}. 
Their particle content includes two \textit{CP}-even bosons (of which one is the Standard Model Higgs
boson), two \textit{CP}-odd bosons (of which one is the pseudoscalar DM
mediator), two charged Higgs bosons, and the \IP. Masses and
couplings of these models are chosen to respect vacuum
stability~\cite{Goncalves:2016iyg} as well as electroweak and flavor
constraints, and to reproduce the observed DM abundance. Figure~\ref{fig:feynman_2}\textit{c} shows an example Feynman diagram.

One can keep going further in this direction. For example, 
{models with multiple mediators} having small couplings
to Standard Model particles have been developed to evade existing LHC
constraints from DM searches~\cite{Duerr:2016tmh}. 

\subsection{Supersymmetric Models and Other Complete Theories}\label{sec:SUSYModels}

So far, we have considered rather general simplified models
inspired by the electroweak sector of the Standard Model.  
Obviously, there are many more possibilities. Additional sources of inspiration for
searches are the large number of BSM theories that have been
developed to solve theoretical problems of the Standard Model as well as the
mechanisms through which these provide {\IP}s.

SUSY is one such class of theories, postulating
partner particles to all Standard Model degrees of freedom. Supersymmetric models 
can stabilize the mass of the light Higgs boson and introduce desired features
in the Standard Model, such as coupling unification. 
Reviews of supersymmetric DM models
can be found elsewhere \cite{Feng:2010gw}. Here, we broadly
sketch models relevant to recent experimental progress and emphasize areas
where we expect future developments.

Supersymmetric DM, the archetype for the concept of weakly interacting massive particles (WIMPs),
has a long history \cite{1984NuPhB.238..453E}. The most
viable and well-studied type of supersymmetric DM has been neutralino DM. The
neutralino, a partner particle to the Standard Model gauge bosons, is
often assumed to be the lightest supersymmetric particle (LSP).
\textit{R} parity conservation makes the LSP stable~\cite{Farrar:1978xj}
and prevents proton decay.

In the MSSM, there are four neutralinos, each of which is a mixture of Standard Model boson
superpartners: a wino, a bino, and two higgsino fermion states.
The lightest neutralino may be called bino-like, wino-like, or
higgsino-like in regions of MSSM parameter space where one of
these components dominates the mixture. 
The phenomenology of the neutralinos is different from that of most of the simplified models described in the previous
section, and it depends on the mixture and on the particle spectrum.
The LHC signatures feature missing transverse momentum from the
neutralino and a high multiplicity of other objects (leptons,
jets) produced in cascade decays of heavier superpartners.

The MSSM is a complete theory with more than 100 independent
parameters, but viable SUSY models might be far simpler. Such
models are used as predictive benchmarks for DM searches. One such model is the phenomenological MSSM (pMSSM), which assumes no
sources of \textit{CP} violation beyond the Standard Model and no flavor-changing neutral
currents, and retains universal couplings and masses for first- and
second-generation superpartners, reducing the number of MSSM
parameters to 19. 

Another DM candidate, found in gauge- or gravity-mediated supersymmetric
models, is the gravitino, a spin-3/2 particle superpartner of the
graviton. Gravitino interactions are suppressed by the Planck
scale (10$^{18}$ GeV) before SUSY breaking. This has consequences
both for the viability as a thermal relic and for the
phenomenology of these models. In gauge-mediated SUSY, the gravitino can be a DM
candidate for a nonstandard cosmological
history~\cite{Dimopoulos:1996vz}. Similar to the neutralino case,
the identity and masses of heavier states decaying to the
gravitino LSP determine the gravitino's phenomenology. However, the gravitino's
interactions are very weak, posing problems for both DD and
ID searches.

Because of the wide variety of potential experimental signatures,
SUSY searches often adopt a simplified model approach,
decoupling the particles that determine the lowest-energy collider
phenomenology [generally the LSP and the next-to-lightest supersymmetric particle (NLSP)] from the rest of the heavier
particle spectrum~\cite{Alves:2011wf}. 
As in the general simplified models described above, extending the MSSM quickly
generates a plethora of nonminimal possibilities.

\subsection{Long-Lived Particle Models}\label{sec:LLPModels}

Another class of models, found within and beyond SUSY, feature suppressed cascade decays of a heavier particle (the NLSP
in SUSY) to a lighter particle (the DM LSP in SUSY). The
suppression can be so large that the particle travels a
macroscopic length within the detector before it decays. Such particles are known as long-lived particles (LLPs).
LHC detectors are not optimized for this purpose, and additional work is required
for searches to be sensitive. 

For example, within SUSY one way to suppress decays is for the NLSP to decay
through a heavy intermediary. Split SUSY models are a
subset of SUSY models in which the gluino must decay through a heavy,
off-shell squark~\cite{Masiero:2004ft}. The heavier the mass of
the squark is, the longer lived the gluino will be. Alternatively, the NLSP
decay can be heavily suppressed by some power of the mass
difference with the LSP. This mass difference also affects DM
coannihilation rates and therefore the DM
abundance~\cite{Ellis:1999mm}. Another way to achieve
long-lived decays is with parametrically small couplings, as in
the case of gauge-mediated SUSY models wherein the
long-lived NLSP decays to its Standard Model partner plus the
gravitino~\cite{Dimopoulos:1996vz}, in a SUSY analog of Cabibbo-suppressed \textit{B} meson decays in the Standard Model. Because of the prevalence of
these mechanisms, it is important to look for long-lived cascade
decays.

In addition to SUSY, small couplings can also lead to find long-lived signatures within the
generic simplified models described in Section \ref{sub:simplifiedModels}. 
The early Universe mechanism that is
responsible for the observed DM density is not known. 
If one assumes thermal freeze-out, then the
coupling of the mediator to DM pairs cannot be arbitrarily small. 
In alternate scenarios,
such as so-called dark freeze-out in which DM can annihilate directly to
BSM mediators but not vice versa, the mediator couplings to the Standard Model
can be much smaller~\cite{Pospelov:2007mp,Das:2010ts,Evans:2017kti}. 
The freeze-in
scenario~\cite{Co:2015pka,Bernal:2017kxu} is another possibility
for reproducing the observed relic density in presence of very weak DM--Standard Model matter interactions,
but LHC rates for some of these models may be too small for observation~\cite{Kahlhoefer:2018xxo}. 

Despite the small couplings, a
sufficiently light mediator can be produced at colliders with
an appreciable cross section. Many such models have been proposed, and here 
we sketch only a few.  
For example, DM can interact with the Standard Model via a dark
vector boson of a \textit{U}(1)$^{\prime}$ dark symmetry, equivalent to the Standard Model's \textit{U}(1)
but with much smaller couplings~\cite{Holdom:1985ag}, such as
those that originate from kinetic mixing. The mediator can also be a
dark scalar boson (a so-called dark Higgs) that couples only to the Standard Model,
akin to a Higgs portal~\cite{Curtin:2014cca}. In both cases, the
dark boson mediator can be light and
long lived~\cite{Pospelov:2007mp}, and its visible decays into Standard Model
particles or associated production with a Standard Model boson provides the
main collider handle for observation~\cite{Curtin:2014cca}. These
scenarios can also be probed by complementary beam-dump and fixed-target experiments~\cite{Battaglieri:2017aum}. Simplified
coannihilation models with long-lived particles have also been
proposed~\cite{ElHedri:2017nny}.

\subsection{Dark Interactions}\label{sec:darkint}

In the above subsections, we have sketched some of the models and signatures
that are currently being sought at colliders. However, we have not covered many other models. 
To an extent, any model containing stable particles interacting
feebly with the Standard Model is a theory of DM.
The key differences between models of DM and other models of BSM
physics are the connections
between these models and astrophysical DM.

The dark sector can be arbitrarily complex, as long as the
particles and interactions it contains satisfy cosmological
observations~\cite{Strassler:2006im, Evans:2017kti}. The models
listed above are simple examples of such dark sectors, where the
mediator particles (e.g., dark bosons) provide the connection with
the Standard Model. Many other models are worthy of mention here, including
asymmetric DM models, in which dark sector particles and
antiparticles are not produced in equal amounts, in the same
fashion as matter and antimatter for Standard Model
baryons~\cite{Zurek:2013wia}; models of neutral
naturalness that realize a mirror copy of the Standard Model without any
low-mass equivalent of the SUSY colored
partners~\cite{Craig:2014aea}; and models of strongly interacting 
massive particles (SIMPs) with to sub-GeV DM candidates~\cite{Hochberg:2014dra}.

\section{EXPERIMENTAL RESULTS}\label{sec:03_ExperimentalResults}

The interactions described in the previous section have many
consequences for astrophysics (where they could modify the DM
density) and for noncollider and collider particle physics
experiments.

Collisions of known particles at high energy, observed with
well-understood detectors, have led to the discovery of many of the fundamental components of
known matter in the Standard Model. While collider experiments alone cannot
discover DM, they can discover {the existence of \IP}s, which could lead the way to
direct study of DM--Standard Model matter mediators in other channels and of
additional particles in a dark sector. 

DM--Standard Model matter interactions may be
feeble because they are mediated by a heavy mediator or by a
mediator with small couplings to Standard Model. Many high-energy collisions are needed to extensively search for these interactions, and the
LHC, which presently collides protons at a center-of-mass energy
of 13~TeV, will deliver both in the coming years.

Below, we outline how the relevant searches are done, some
of their challenges, and the information the searches can provide about the properties
of hypothetical particles (couplings, mediator mass, other
parameters of the Lagrangian in a particular model). In Section~\ref{sec:04_Extrapolation}, 
we describe how collider information can be related to
noncollider DM searches and to the present DM abundance.

\subsection{Searches for Invisible-Particle Production Mediated by Standard Model Bosons}\label{sec:results_ZHSearches}

Colliders have already provided spectacular evidence for copious production of low-mass invisible particles: the huge rate of neutrino production
mediated by the \textit{W} and \textit{Z} bosons. As a result, neutrino production via the \textit{Z}
boson is often the largest background to searches for new {\IP}s
and is important to understand well. 
The rate of events with considerable \pt is predicted by the Standard Model.
The data would show a significant deviation from this prediction if 
the \textit{Z} boson were coupled to additional {\IP}s
lighter than approximately half its mass. 
The most precise measurement of
the invisible \textit{Z} boson's width, 499.1 $\pm$ 1.5 MeV, has been inferred from
the total \textit{Z}
width at LEP~\cite{ALEPH:2005ab}. This value can be used to
constrain the parameters of models such as \textit{Z}
portals~\cite{Carena:2003aj,Escudero:2016gzx}, where the coupling
between the \textit{Z} boson and an invisible Dirac fermion, 
lighter than the \textit{Z} mass, is constrained to be
significantly smaller than the values necessary for a thermal relic.
A less precise direct measurement of the \textit{Z} boson's
invisible width, also by LEP, uses invisible decays with a photon
emitted as initial-state radiation (ISR), selecting events with a
single photon, the total missing transverse momentum inferred
from momentum balance with the visible particles (\MET), and
little other event activity.  At the
LHC, precision measurements continue to test the production and
decay of \textit{Z} bosons for the effects of {\IP}s. 
For example, ATLAS
has measured the ratio of cross sections for jet and \MET
production, dominated by invisibly decaying \textit{Z} bosons, to
the production of \textit{Z} bosons decaying to dilepton pairs, a ratio that
is sensitive to the production of additional
{\IP}s~\cite{Aaboud:2017buf}.

Invisible decays of the newly discovered Higgs boson are, in the Standard Model,
decays to a pair of \textit{Z} bosons that then each decay invisibly,
contributing to less than 0.1\% of the total width of the Higgs boson. 
With present data, the Higgs-to-invisible rate could become observable
if the Higgs is coupled to additional
{\IP}s~\cite{Khachatryan:2016vau,Aad:2015pla}. To constrain the
invisible width of the Higgs, ATLAS and CMS cannot directly
measure its total width in a model-independent
fashion~\cite{Dobrescu:2012td}; instead, searches attempt to
directly observe these decays via their recoil against visible
particles (resulting in substantial \MET) or through a comparison of measurements of the
Higgs parameters under additional assumptions about the BSM
physics. Direct Higgs-to-invisible searches have used Run 1 and
Run 2 data, combining several strong and electroweak production
channels. A combination of DD and ID searches yields the most
stringent upper bound on the fraction of invisible decays of the Higgs
boson: 23\%~\cite{Khachatryan:2016whc,Aad:2015pla}. 
For Dirac {\IP}s much lighter than half the Higgs mass, 
this places constraints on Higgs portal couplings that are smaller than those necessary for a thermal relic, 
indicating that, even if this model with these parameters is realized in nature, 
additional sources of DM are needed. 

\subsection{Generic Searches for Invisible Particles from Beyond-the-Standard-Model Mediation}\label{sec:results_monoXSearches}

Searches for invisible decays via a Standard Model mediator (the \textit{Z} or Higgs boson)
can be viewed as special cases of searches for more general BSM
mediation of {\IP}s. Mediator decays to {\IP}s are
suppressed if the {invisible-particle} mass is heavier than half the mediator mass.
For the case of the $Z$ or $H$ boson--mediated interactions, the upper
bound for {\IP} masses that can be observed is $\sim$45--65 GeV. 
Moreover, the distribution of \MET in events with a mediator with masses comparable 
to the $H$ and $Z$ has a similar shape to that of
the \textit{Z} boson--mediated neutrino background.

\begin{textbox}
\section{MEASURING INVISIBLE PARTICLES: \MET RECONSTRUCTION}

\noindent Measurements of \MET in experiments at hadron colliders 
should include contributions from all particles in the event, 
and therefore rely on precise measurements in all detector systems. 
The calculation of \MET includes all visible physics objects 
(e.g., jets, leptons reconstructed from energy deposits and tracks)
from  the hard scatter interaction. 
Contributions that are not attributed to physics objects form the
soft component of the \MET~\cite{Aad:2016nrq,CMS-PAS-JME-16-004}.

A challenge for \MET measurements is to exclude contributions from
the debris of additional proton--proton interactions
detected at nearly the same time as the hard scatter (pileup). 
The combination of tracking and calorimeter information is used to identify
tracks and energy deposits that originate from the primary collision
vertex~\cite{CMS-PAS-JME-16-004,ATLAS-CONF-2014-019}. 

A further challenge for the measurement of invisible particles is
the rejection of fake missing transverse momentum. 
Noncollision backgrounds, such as cosmic rays, beam background, and detector
noise, make a significant contribution to the tails of the \MET
spectrum (Figure~\ref{fig:fakeMET}). Specific quality
cuts, based on the presence of tracks associated with the deposited
energy and on the energy deposited in the various calorimeter layers,
are applied to reject these events~\cite{ATLAS-CONF-2015-029}. For example, the
number of events passing the jet+\MET analysis selection before
these quality cuts is approximately 10 times larger than the Standard Model
contribution in Reference~\citen{Aaboud:2016tnv}.
\end{textbox}

For heavier BSM mediators, this is not necessarily the case. Their
decay to {\IP}s can produce \MET distributions substantially
different from the Standard Model background. The complexity of the processes
mediating invisible-particle production determines the composition
of the visible recoil, so searches are employed across many
different visible-particle signatures. 
Because of the large number of model possibilities, many collider
searches, from LEP to the Tevatron to the most recent LHC searches
(e.g.,~\citen{Fox:2011fx,Bai:2010hh}), aim to be model agnostic,
designed to detect an excess of \MET over the Standard Model background with
minimal assumptions about the visible objects in the recoil.
For this reason, the ISR+\MET has become a key
signature for invisible-particle searches at colliders
and has gained popularity since its use at LEP~\cite{Birkedal:2004xn}. 

We begin with the jet+\MET search, which illustrates
techniques used in other general invisible-particle searches and
shares with them many of the same challenges in measuring \MET
(some of which are outlined in the sidebar titled Measuring Invisible Particles: \MET\ Reconstruction).
Traditionally, these
have been called mono-\textit{X} searches, but the radiation of a single
object is only the leading process in the simplest
reactions~\cite{Haisch:2013ata}.

Only a fraction of proton--proton collision events can be recorded for
further processing. The selection of those events (termed triggering; see Reference~\citen{Smith:2016vcs}
for a review of the ATLAS and CMS systems and Reference \citen{Aaij:2012me} for LHCb)
needs to happen in real time. This selection can be done in a model-agnostic way by looking
for substantial \MET; however, for models that do not
produce large \MET, one is forced to assume more about the visible
recoil, as described in the rest of this section. 

\subsubsection{Searches with jets}\label{sub:monojet}

One way to reduce the model dependence of DM searches at colliders is to
require that the recoiling visible particles be governed by Standard Model
processes, not by the dark interaction, so their relative rates and spectra are predictable
with no model assumptions. ISR meets this criteria.
Standard Model bosons are likely to be present in any BSM process, radiated
from initial-state partons at rates fixed by the Standard Model. Because gluon
ISR is far more prevalent at hadron colliders than the other forms, the jet+\MET
search is key in this approach.

LHC jet+\MET searches~\cite{Aaboud:2017phn,Sirunyan:2017jix}
typically selects collision events with a moderate amount of \MET
(above roughly 200 GeV in the 13 TeV analyses in order to trigger at a manageable rate;
see the sidebar titled Challenges for Triggering Low-Mass\ Resonances at Hadron Colliders in Section~\ref{sub:visibledecays})
and at least one jet with \pt higher than 100--200 GeV in the central
region of the detector (with pseudorapidity $|\eta|<2.4$).
From this sample, further
restrictions on additional hadronic jets and other visible
particles are used to suppress contributions from Standard Model processes and
from instrumental backgrounds causing spurious \MET. These
requirements reduce the generality of the analysis but also better
isolate signal-like events. For example, contributions from \textit{W}
bosons decaying to leptons are reduced by vetoing events with
leptons, and top quark pair production is reduced by limiting the number
of jets present. The remaining Standard Model multijet processes can exhibit
high \MET when one or more jets are mismeasured. Such
mismeasurements often result in \MET along the axis of a jet, and
this feature is used to reduce the background to approximately 1\% of the
total. Noncollision events (e.g., intersecting cosmic rays,
beam--gas interactions, and calorimeter problems) can also produce
spurious \MET. Figure~\ref{fig:fakeMET} shows that such
events dominate a high-\MET data sample unless they are rejected with
criteria tailored to the expected collision time and detector
hardware.

\begin{figure}[!htpb]
\includegraphics[width=0.5\textwidth]{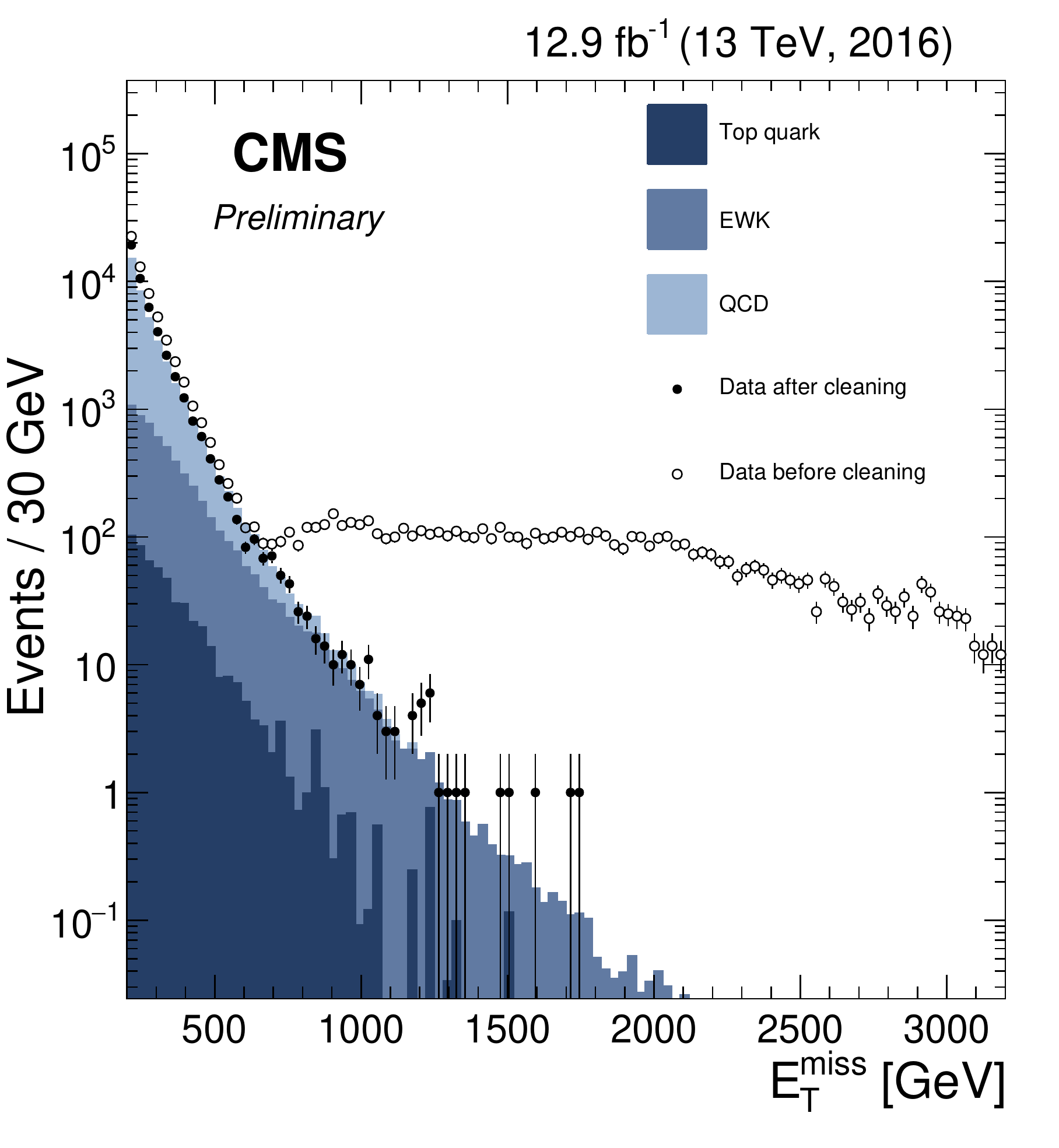}
\caption{The \MET distribution of events, termed as $E_T^{miss}$ in the x axis, selected for high total
hadronic energy and at least two jets with \pt{} $>$ 400 and 200
GeV, before (\textit{open circles}) and after (\textit{filled circles}) rejection of
spurious \MET backgrounds~\cite{CMS-PAS-JME-16-004}. The
predictions of Monte Carlo  simulations (\textit{shaded areas}) are also shown. Strong
noncollision background suppression is vital to \textit{X}+\MET analyses.}
\label{fig:fakeMET}
\end{figure}

After the above criteria are satisfied, one arrives at a sample composed mainly
of invisible decays of the \textit{Z} boson (approximately 55--70\% of the
total background). A substantial rate of semileptonic decays of
the \textit{W} boson also survives the lepton veto when the lepton is not
reconstructed (approximately 20--35\% of the total background).
The main observable is typically the number of events in one or
more \MET regions with events selected to be enriched in contributions from signal processes (signal regions). 
Signal regions are either exclusive (in bins of
\MET) or inclusive (considering all events above a given \MET
threshold).

Because {\IP}s have feeble interactions with the colliding
partons, and thus low production cross sections, these searches
need precise estimates of the shapes of the backgrounds, especially in the low-\MET regions.
A background estimate made solely on the basis of Monte Carlo simulation is subject to
uncertainties in both theory and detector simulation affecting the
total cross sections, and therefore is not precise enough. 
Recent ATLAS and CMS searches combine the information from data in signal-free
{control regions} selecting visible-boson (\textit{W},
\textit{Z}, $\gamma$)+jet processes with the most recent perturbative
calculations~\cite{Lindert:2017olm}, to
estimate the \textit{Z}- and \textit{W}-mediated neutrino backgrounds more precisely.
ATLAS\ estimates backgrounds from top processes using a
control region with $b$ jets, while CMS takes this background from
simulation. Estimates of smaller backgrounds rely more heavily on
simulation.

Currently, the precision achieved for the background estimate is 2--7\%
(CMS) and 2--10\% (ATLAS), depending on the \MET range. The
remaining uncertainties arise mainly from the identification of
leptons (CMS) and the understanding of jet and \MET
calibration (ATLAS).
With no excesses observed, these searches set 95\%-CL limits
on the production cross section of {\IP}s, typically ranging from
0.5 \pb to 2 \fb, depending on the \MET threshold. 

ATLAS and CMS report
constraints for a selection of mediator models and parameters. These constraints are strong enough to probe (axial--)vector mediated
processes, but searches are only becoming sensitive to lower cross-section scalar mediated processes. 
 These constraints
can be interpreted as limits on the interactions between the
mediator and the Standard Model (e.g., \gq) under specific sets of model
assumptions, not on the mass and other properties of the {\IP}s
per se. As an example, for the simplified model with
(axial--)vector mediators, mediator masses of up to
1.5--1.9~TeV are ruled out
for an invisible coupling of \gdm = 1 and \gq = 0.25. For mediators lighter than
this bound, the search can exclude Standard Model couplings of order 0.1 or,
alternatively, lower \gdm values than unity. With this amount of
data, the searches are also becoming sensitive to lower-rate
interactions mediated by scalar mediators, and the
ATLAS search~\cite{Aaboud:2017phn} sets explicit constraints on
colored scalar mediators, where, for unit couplings and invisible
particle masses of up to 100 GeV, the mass of the mediator is
constrained to be above 1.7 TeV. Jet+\MET results from LHC Run 1
and the Tevatron have also reported constraints on EFT models.

Since this type of search can constrain a wider variety of
interactions than explicitly considered, steps have been taken to
allow easy reinterpretation of the results. ATLAS and CMS provide
more detailed experimental results on the HEPData
platform~\cite{Maguire:2017ypu}. CMS also provides a simplified
likelihood function encapsulating the
result~\cite{Collaboration:2242860,Sirunyan:2017jix}.

\subsubsection{Searches with photons and vector bosons}\label{subsub:monoV}

Particles other than gluons can constitute visible-particle
recoil. In models where the recoil arises from ISR, the rates
for photon and electroweak boson radiation are much smaller than
for gluon radiation. Nevertheless, searches in photon+\MET and
\textit{Z}+\MET channels can play a complementary role alongside jet+\MET\ searches,
with a smaller and different mix of backgrounds and different
systematic uncertainties. Both ATLAS and CMS have performed
searches in each channel. With lower backgrounds, events can be
recorded with lower kinematic thresholds, resulting in lower \MET
and visible-\pt{} selections. For example, the lowest \MET value
probed by the \textit{Z}+\MET search, where the \textit{Z} decays into
leptons~\cite{Sirunyan:2017qfc,Aaboud:2017bja}, is around 100 GeV,
versus 200 GeV for the jet+\MET searches (e.g.~\cite{Sirunyan:2017jix}).

These searches can play a much more powerful role when the recoil
arises from the dark interaction itself rather than ISR. In these
cases, photon or vector boson recoil
(e.g.,~\citen{Birkedal:2004xn,Gershtein:2008bf,Petriello:2008pu}), rather than gluon
recoil, may be the dominant signature. The
event selection and the background estimation strategies generally
mirror those of the jet+\MET search, but vary with the type of
recoil, taking advantage of the special features of the signal. 
Photon+\MET searches~\cite{Aaboud:2017dor,CMS-PAS-EXO-16-053} has lower backgrounds
with respect to the jet+\MET searches and therefore retains a comparable albeit
lower sensitivity. 

Some searches look for more complex recoil features. For example, 
the searches for signatures of hadronic decays
of high-\pT{} electroweak bosons recoiling against sizable \MET
take advantage of the boson boost to reconstruct its collimated decay products
in a single, large-radius jet~\cite{Sirunyan:2017jix,ATLAS-CONF-2018-005}. 
Vector boson jets have a typical two-prong pattern from the hadronization
of the quark--antiquark pair, while QCD jets do not present any structure.
Substructure techniques (see Reference~\citen{Larkoski:2017jix} for a review) are
used to discriminate between these two cases. 

None of the photon and vector boson+\MET searches have yet observed a signal. These searches are generally not
as sensitive as the jet+\MET search to models in which the
visible recoil arises from ISR, because of smaller signal
acceptance and a comparable signal-to-background ratio. Nevertheless, they can
be remarkably competitive:\ The
photon+\MET searches are the next-most-powerful probe after the jet+\MET searches. Moreover,
these searches provide the most stringent limits on some models
in which the boson in question is directly involved in the dark
interaction~\cite{Berlin:2014cfa}.

\subsubsection{Search signatures including the Higgs boson}

One can also look for the newly discovered Higgs boson in the
recoil. Due to the heavy mass of the Higgs and the small
heavy-flavor content of the proton, the rate of Higgs ISR is
insignificant. Thus, searches for Higgs+\MET target dark
interactions in which the Higgs is a direct participant and,
therefore, the interaction is closely tied to the Higgs sector.
This is a feature of many models that extend the Standard Model scalar sector, 
such as those described in Sections~\ref{sub:simplifiedModels} and~\ref{sec:LessSimplifiedModels}. 

Dedicated searches for Higgs+\MET  select Higgs events similarly to
the inclusive Higgs measurements, then require substantial \MET to
reduce the backgrounds to the search. In the Run 2 data, searches
in the 
$H \rightarrow \gamma\gamma$~\cite{Sirunyan:2018fpy,Aaboud:2017uak}, 
$H \rightarrow \tau\tau$~\cite{Sirunyan:2018fpy}, 
and
$H \rightarrow b\bar{b}$~\cite{Aaboud:2017yqz,CMS-PAS-EXO-16-050} channels have been
performed. Searches in the $ZZ$ and~$WW$ channels are
expected to contribute as well, once substantially more data have been
collected.

The \MET+$H \rightarrow \gamma\gamma$
searches~\cite{Sirunyan:2018fpy,Aaboud:2017uak} benefit from
their ability to precisely constrain the diphoton pair to the
Higgs boson mass. They are still statistically limited. The
relatively low backgrounds enable probing for anomalous \MET as low
as 50 GeV~\cite{Sirunyan:2018fpy}. The diphoton-invariant mass
is fitted in different signal categories, each optimized for
different types of signal models. 
The $H \rightarrow \tau\tau$ search also benefits from lower SM backgrounds 
than other \MET+$H$ searches, as it selects events based on the presence 
of $\tau$ leptons rather than on the presence of large \MET. This search 
is sensitive to anomalous \MET signals above 105 GeV. 
The search for a Higgs boson decaying to
two bottom quarks~\cite{Aaboud:2017yqz,CMS-PAS-EXO-16-050} requires \MET$>$150 or 200 GeV.
These searches employ jet substructure techniques to
select boosted Higgs decays amid a background of QCD processes.
The majority of the backgrounds are estimated using
data-driven techniques and Monte Carlo simulation and constrained in dedicated control regions. 

In the absence of a signal, limits are placed on the baryonic Higgs benchmark
model (outlined in Section~\ref{sub:simplifiedModels} and shown in Figure~\ref{fig:feynman_1}\textit{a})
with \gq = 1/3 (ATLAS) or 0.25 (CMS), \gdm = 1, and \ghZprimeZprime/$m_{Z}$ = 1, on a
\Zprime--2HDM (Figure~\ref{fig:feynman_1}\textit{b}).
\footnote{In the case of the \Zprime--2HDM, CMS and ATLAS set different masses for the new Higgs
bosons, so the constraints are not yet directly comparable.} 
Both Higgs+\MET and \textit{Z}+\MET searches are also sensitive to extended
scalar sectors such as two Higgs doublets with a scalar or
pseudoscalar mediator~\cite{Bauer:2017ota,Goncalves:2016iyg,Bell:2016ekl};
the search in Reference~\cite{CMS-PAS-EXO-16-050} sets limits on the 2HDM + pseudoscalar model of~\cite{Bauer:2017ota}
shown in Figure~\ref{fig:feynman_1}\textit{c}.

\subsubsection{Searches with third-generation quarks}

In scalar- and pseudoscalar-mediated simplified models, 
the mediator can be produced along with two top or
bottom quarks, leading to a signature that includes \MET and multiple \textit{b} jets.
A recent ATLAS search in these channels~\cite{Aaboud:2017rzf} is
optimized for both recoil consisting of semileptonic and fully
hadronic top quark decays and recoil with one or two bottom
quarks. This signature is similar to that of third-generation
quark superpartners and can be part of dedicated SUSY searches or
used for reinterpretation~\cite{Aaboud:2017aeu,Sirunyan:2017leh}.
SUSY searches suppress most of the $t\bar{t}$ background, matching
specific models to specific, low-background signal regions.
Relative to these approaches, the search described in
Reference~\citen{Aaboud:2017rzf} is less narrowly targeted at specific
models, where control regions can be more reliably developed,
instead relying more heavily on simulation. The sensitivity of
searches of \MET associated with top quarks is comparable for the
two strategies.

No significant excess is observed in these searches. For invisible-particle masses of
1 GeV, color-neutral pseudoscalar mediators with masses in the range of 20--50
GeV~\cite{Aaboud:2017aeu} and scalar mediators with masses up to 100
GeV~\cite{Sirunyan:2017leh} are excluded. Signatures with
$b\bar{b}$ pairs are less sensitive to models that do not
explicitly privilege bottom quarks, but they can set much higher limits
on colored mediator masses in the case of preferential couplings to
bottom quarks~\cite{Agrawal:2014una}.

Other LHC searches in this category are those including only
one top or bottom quark (also called monotop or monobottom
searches), e.g.~\cite{Sirunyan:2018gka, Aad:2014wza}. They place
constraints on models that include singly produced {\IP}s through
flavor-changing neutral currents~\cite{Boucheneb:2014wza}.

\subsection{Searches for Supersymmetric Invisible Particles}\label{sec:results_SUSYSearches}

So far motivated by simple models, the \textit{X}+\MET experimental
searches discussed make few choices about the visible recoil
particles (i.e., the species of a single particle), yet they have
already led to a plethora of diverse signatures. Models with more degrees of freedom vastly expand the set of
signatures to be explored. SUSY adds, along with
{\IP}s, a full copy of the Standard Model particle
spectrum to be discovered. Each superpartner features particular decay chains that
can be targeted to a greater or lesser degree, privileging either
generality or maximal sensitivity. Compared with the 
searches described above, SUSY searches generally opt for a more specific
decay topology and thus apply more stringent event selections
based on the expected kinematic features, often using
discriminating variables based on the combined mass of visible and
invisible particles (e.g.,~\citen{Lester:1999tx}) to recover
the ability to resolve various resonant decays.

SUSY has received much attention from ATLAS and CMS, and many
searches for its particles, such as for squarks and gluinos, have
a long history at earlier colliders as well. In this section, we
discuss experimental results with explicit connections to DM.

So far, no SUSY search has produced a conclusive signal. However,
given its multifarious signatures, it is difficult to make general
statements about the current status, even for simplified models of
SUSY. Perhaps the best one can say is that searches for
strongly produced superpartners constrain them for masses
approaching $\sim$2 TeV for neutralino masses up to 1 TeV
(e.g.,~\citen{Aaboud:2017bac,Sirunyan:2017yse}); by contrast, other processes are
less constrained. Direct production of weakly coupled
superpartners has a much smaller production rate; therefore, the
constraints on them are significantly weaker, as shown in Figure~\ref{fig:SUSYSummary_ew}.
Third-generation squarks are generally only constrained to be at least several
hundred GeV for neutralinos of similar masses (e.g., \citen{Sirunyan:2017wif,Aaboud:2016wna}). But there are
numerous exceptions to these blanket statements, even before one
takes into account that the masses exclusions shown in Figure~\ref{fig:SUSYSummary_ew} 
apply only to specific slices of a multidimensional model parameter
space.

\begin{figure}[!htpb]
\includegraphics[width=0.7\textwidth]{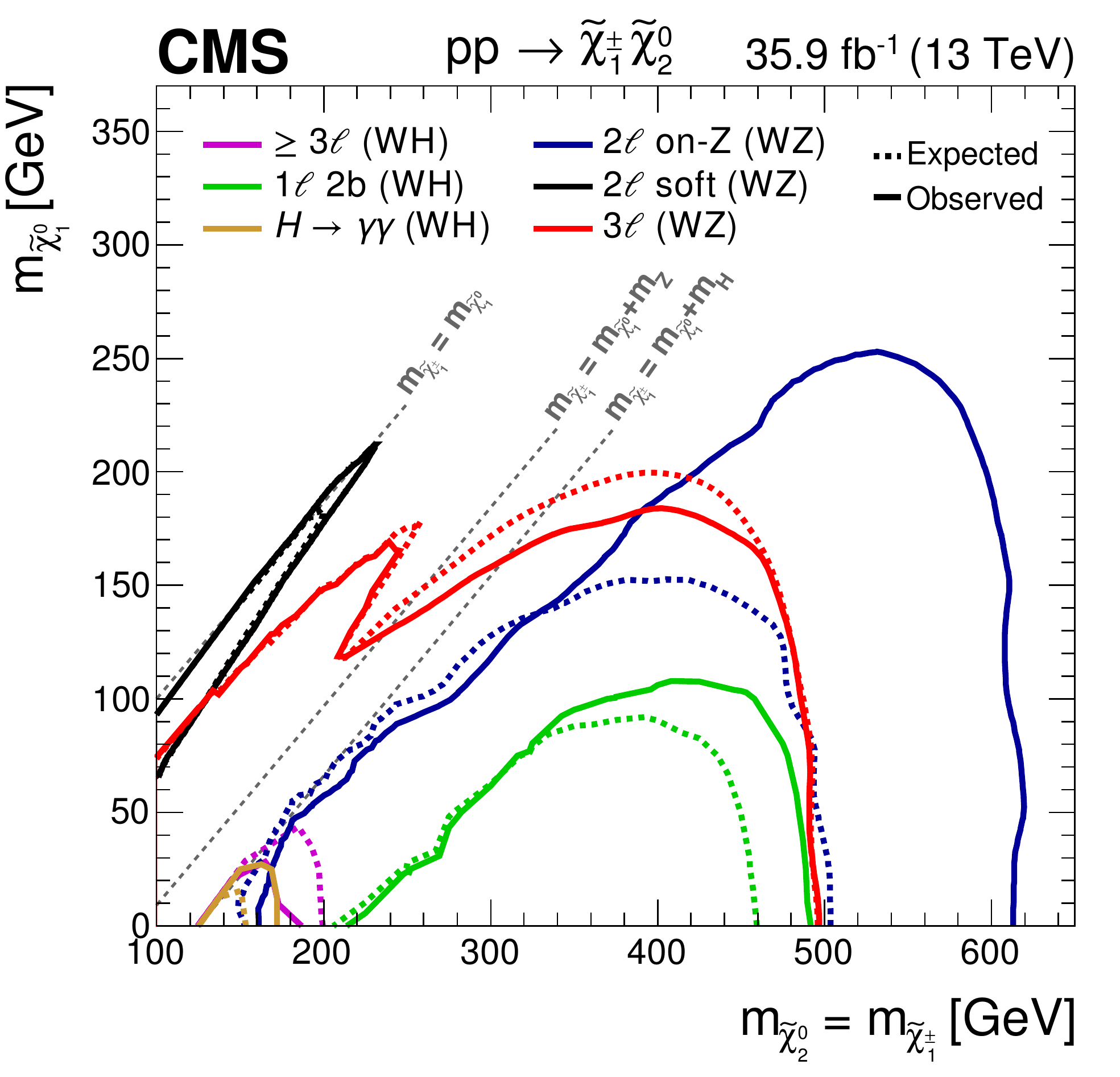}
\includegraphics[width=0.7\textwidth]{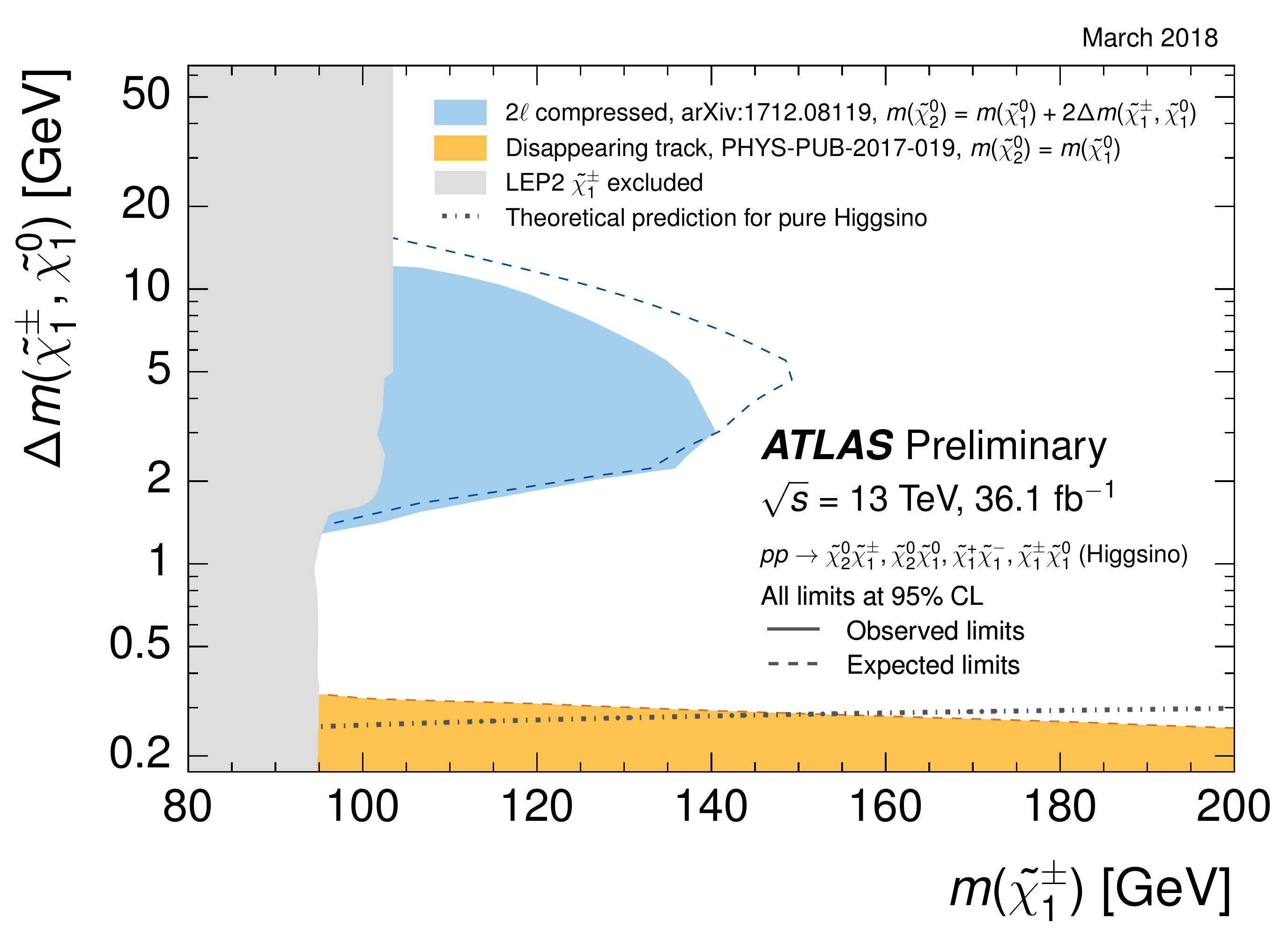}
\caption{Mass reach of (\textit{a})\ CMS and (\textit{b})\ ATLAS searches for a selection of
results targeting electroweak supersymmetry production, available as of
July 2018.
Panel \textit{a} adapted from Reference \citen{Sirunyan:2018ubx}. 
Panel \textit{b} adapted from Reference~\citen{ATL-PHYS-PUB-2017-019}.\label{fig:SUSYSummary_ew}}
\end{figure}

Although LHC searches have so far probed mainly  SUSY channels that are strongly produced, searches for more rare processes are now entering their prime. With the data now collected, one can explore the
electroweakino parameter space (e.g.,~\citen{Sirunyan:2018ubx,ATLAS:2017uun}). Searches for gauge
boson superpartners (gauginos) can reach approximately 1 TeV if
the superpartners of Standard Model leptons are light, and the search can
benefit from a high leptonic branching ratio, whereas their reach
is lower if their decays proceed through \textit{W} and \textit{Z} bosons. More
luminosity also provides access to new regions of parameter space
for specific signatures, such as so-called compressed regions where small
mass differences between superpartners cause the signals to lie
buried in large backgrounds at low
\MET~\cite{Aaboud:2017leg,Sirunyan:2017zss}. Small mass
differences can also suppress superpartner decays, resulting in
long lifetimes that can be exploited to study regions with a mass
difference as low as 0.2 GeV for higgsino models~\cite{ATL-PHYS-PUB-2017-019}. Other mechanisms of decay
suppression can do this as well [e.g., split
SUSY~\cite{Sirunyan:2018vjp}].

Despite the unwieldy diversity SUSY signatures, a sufficiently
specific model can provide a concrete framework on which to build
an understanding of the combined effect of many experimental
constraints. The study reported in Reference \citen{Conley:2010du}, continued by the LHC
experimental collaborations~\cite{Aad:2015baa, Khachatryan:2016nvf}, uses the
pMSSM to define a finite (although large) parameter space for
which the wealth of experimental constraints can be systematically
evaluated, and underexamined signatures can be considered for future emphasis.
One may also identify the LSP with astrophysical DM
to focus more specifically on regions compatible with a given
cosmology~(e.g., \citen{Aaboud:2016wna}).

Collaborations such as GAMBIT~\cite{Athron:2017ard} and
Mastercode~\cite{Bagnaschi:2017tru} have combined a variety of
tools to aid in such efforts. These codes encapsulate search
results in statistical outputs that can be combined to construct
global constraints for the models of interest. For example, one
may compile the likelihood functions for the parameters of a SUSY
model given the results from collider, DD, and ID experiments.

\subsection{Searches for Long-Lived Particles}\label{sec:results_LLPSearches}

Prompt decays produce a visible recoil that originates at the collision
point; thus, it can be reconstructed using the techniques for
which the experiments were designed. The long-lived mediators
described in Section~\ref{sec:LLPModels} present different
experimental challenges. For short lifetimes, LLPs can decay inside the tracking detectors, appearing as
displaced decay vertices. Some SUSY decay chains lead to
disappearing tracks, if the visible particles decay into the LSP
and soft particles (e.g.,~\citen{Aaboud:2017mpt, CMS:2014gxa}).
Even longer-lived particles can decay in the calorimeters or in
the muon spectrometers, or they may exit the detector cavern
completely before decaying. 
These signatures add yet another
dimension of complexity to such searches, because observing these
events may require dedicated triggers, reconstruction algorithms,
and even detectors \cite{Ball:2016zrp,Chou:2016lxi}.

Searches at colliders use a variety of experimental signatures to
target different types of dark bosons, such as dark vector or
scalar bosons. An LHCb search for dimuon
resonances~\cite{Aaij:2017rft} is sensitive to visible decays of
vector mediators in the mass range between 10 and 70 GeV. This
search can use the entire sample of dimuon decays delivered to
LHCb, recorded at the full collision rate directly at the trigger
level~\cite{Aaij:2016rxn}, also placing constraints on dark bosons with longer lifetimes. 
Below 10 GeV, experiments at
electron--positron colliders have searched for dilepton resonances
or missing mass produced in association with ISR photons (e.g.,~\citen{Lees:2014xha,Lees:2017lec}). 
LHCb also searches dimuon
events for scalar bosons with masses between 250 MeV and 4.7
GeV~\cite{Aaij:2016qsm}, for a range of lifetimes. Dark bosons can
also arise in Higgs decays via a hidden-sector mechanism. For
example, the searches described in References \citen{ATLAS:2016jza} and \citen{CMS-PAS-HIG-16-035}
look for exotic Higgs decays into collimated lepton jets,
constraining the decay rate to be below 10\% for a range of dark
photon lifetimes. Reference~\citen{Curtin:2014cca} provides a review of this and other
possible benchmark models.

\subsection{Consequences of Neutral-Mediated Models: Visible Decays}\label{sec:MediatorSearches} \label{sub:twoBody}
\label{sub:visibledecays}

\begin{textbox}
\section{CHALLENGES FOR TRIGGERING LOW-MASS RESONANCES AT HADRON COLLIDERS}

\noindent The LHC collides protons every 25 $\mathrm{ns}$ in nominal
conditions. The decision to record collision events for further
analysis is made by each experiment's trigger
system~\cite{Smith:2016vcs,Aaboud:2016leb,Khachatryan:2016bia,Aaij:2012me}.
Its first, hardware-based level uses partial detector information
for fast decisions. Its second, software-based level uses more
refined algorithms and has access to further detector information.

A challenge for many DM searches at colliders is to trigger on events with low-\pt objects.  
The trigger system records events above a certain threshold (e.g., leading-jet \pt or event
\MET), since energetic processes are likely to contain interesting
features. Only a small fraction of events below these thresholds
is recorded, penalizing signals with lower-energy signatures.
However, if only final-state objects reconstructed by the trigger
system are recorded, instead of full event information, the
storage limitations can be
overcome~\cite{Aaij:2016rxn,Khachatryan:2016ecr,Aaboud:2018fzt}.
Alternative strategies to access resonances with masses below the TeV 
are to trigger on the ISR or selecting lower-backgrounds heavy quarks in the final state,
as described in the text. 

Pileup can add energy uncorrelated to the hard process of interest, increasing the event
rate for a given trigger threshold; trigger \MET rates increase
exponentially with the number of additional interactions. For this
reason, increases in LHC instantaneous luminosity and data set size
come at the cost of increased thresholds. Dedicated pileup
suppression algorithms, including partial tracking information, are
used in trigger
reconstruction~\cite{CMS:2014ata,ATLAS-CONF-2014-019}. ATLAS and
CMS foresee dedicated hardware systems to obtain full tracking
information in future LHC
runs~\cite{Shochet:2013gaw,1748-0221-6-12-C12065}.
\end{textbox}

Dark interactions might also be probed without actually producing
{\IP}s. For example, if the mediator particle can be produced via
interactions with quarks, it may also decay into quarks. In this
case, it may be discovered in dijet, di--\textit{b} jet, and ditop resonance searches
(e.g.,~\citen{Liew:2016oon,Chala:2015ama}).

Dijet resonance searches have been routinely used at hadron
colliders to probe for heavy particles at newly reached collision
energies. They exploit an expected absence of features in the
dijet-invariant mass distribution to estimate the search
background directly from a fit to the data, minimizing modeling
and theory uncertainties. This permits the observation of low-rate
localized excesses (width/mass of up to $\sim$15\% and $\sim$30\%
for ATLAS and CMS respectively) from resonant
dijet production~\cite{Aaboud:2017yvp,Sirunyan:2018xlo}. For
wider signals, searches exploiting the scattering angle of dijet
events can be used~\cite{CMS-PAS-EXO-16-046,Aaboud:2017yvp}.

At the LHC, typical dijet searches lose sensitivity at masses
below approximately 1 TeV~\cite{An:2012ue,Dobrescu:2013coa}, where high
rates force the experiments to discard a large fraction of the
data at the level of the trigger (see the sidebar titled Challenges for Triggering Low-Mass Resonances at Hadron Colliders). 
However, by recording much less
information for these low-mass
events~\cite{Khachatryan:2016ecr,Aaboud:2016leb}, 
one can reduce this threshold to values as low as 450
GeV~\cite{Sirunyan:2018xlo,Aaboud:2018fzt}.  
Alternatively, one can look at the subset of dijet events in which a
high-\pt{} ISR object happens to
trigger~\cite{ATLAS:2016bvn,Sirunyan:2017nvi}, in a similar fashion as in
the jet+\MET searches described in Section~\ref{sub:monojet}. 
LHC searches are sensitive to even lower mediator masses when
reconstructing the mediator decay products into a single jet and
employing substructure techniques~\cite{Sirunyan:2017nvi,Aaboud:2018zba}. 
Dedicated searches for resonances of third-generation
quarks, with lower backgrounds and therefore lower thresholds, are
also performed~\cite{Aaboud:2018tqo,CMS-PAS-HIG-16-025,Aaboud:2017hnm}.

Resonance searches can also constrain mediator couplings to 
leptons~\cite{Aaboud:2017buh,Khachatryan:2016zqb}. For dielectron
and dimuon searches, the main backgrounds arise from Drell--Yan
processes, which are estimated with simulations corrected for next-to-next-to-leading-order
effects and normalized to the \textit{Z} boson yield in the data.

\begin{figure}[!htpb]
\includegraphics[width=\textwidth]{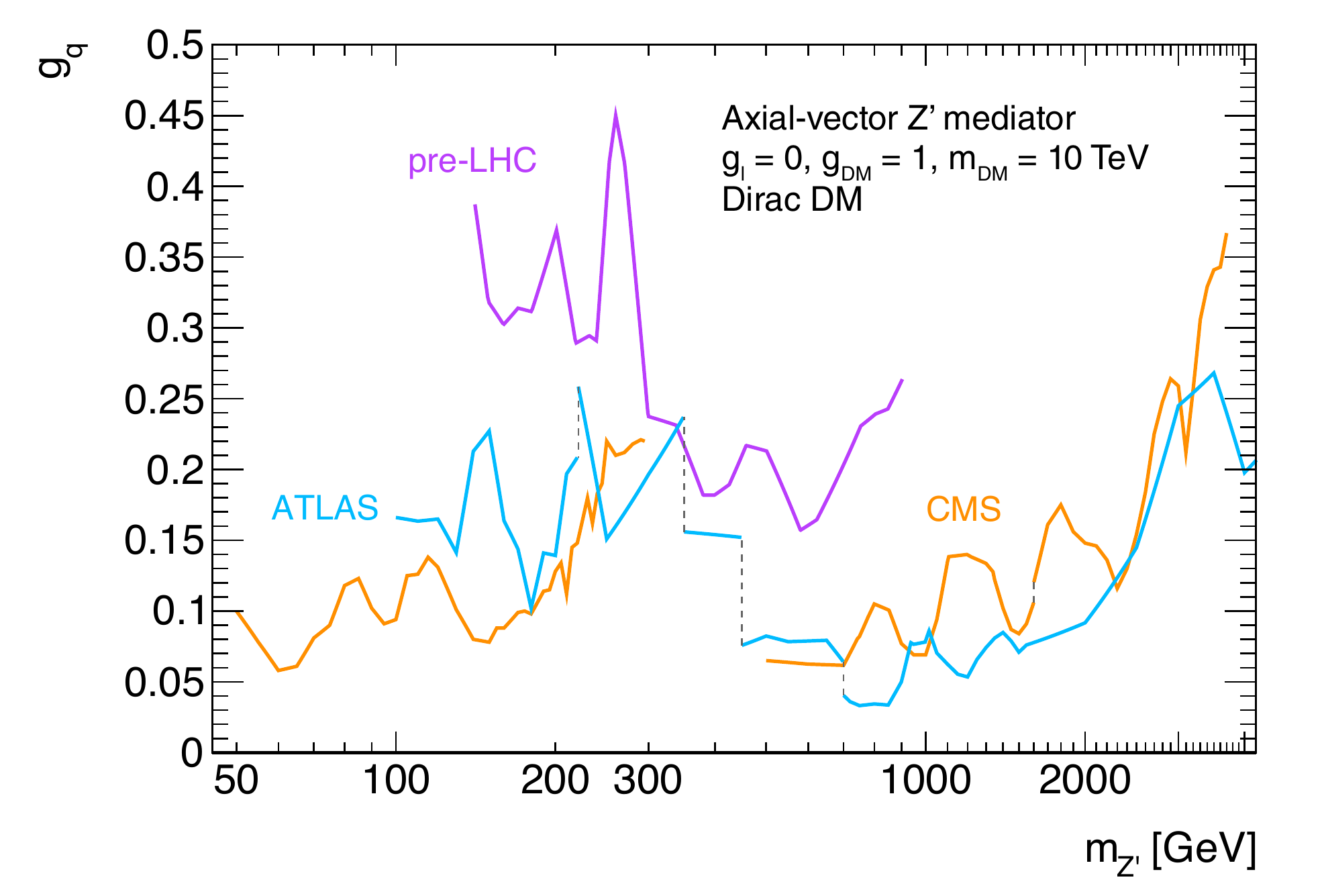}
\caption{Summary of constraints from searches for narrow, light dijet
resonances from ATLAS and CMS available as of
July 2018, where discrete points are taken
from the coupling-mass limits (as the best limit for each of the points)
on a simplified model mediated by an
axial--vector $Z^\prime$ coupling exclusively to quarks from the
searches mentioned in the text, and interpolated at the crossings.
The coupling to leptons ($g_l$) is set to zero and the coupling to 
DM ($g_{DM}$, also termed as $g_{\chi}$ in the text) is set to unity. For this
comparison of dijet searches, the \Zprime is not allowed to decay on-shell
to DM particles since the mass of the DM particle is set to 10 TeV. 
Couplings above the lines are excluded at 95\% CL, up to the
values where larger couplings yield a resonance width larger than
10-15\% (roughly \gq $>$ 0.5). Abbreviation:\ DM, dark matter.
Pre-LHC constraints are extracted from Reference \citen{Dobrescu:2013coa},
while LHC constraints are taken from References \cite{Khachatryan:2016ecr,
Khachatryan:2015sja,ATLAS:2016bvn,Sirunyan:2017nvi,Aaboud:2018zba,
Sirunyan:2017nvi,Sirunyan:2018xlo,Aaboud:2018fzt,Aaboud:2017yvp}.
}
\label{fig:couplingmass}
\end{figure}

Figure \ref{fig:couplingmass} illustrates constraints from the
dijet resonance searches mentioned above, on the quark coupling of
the mediator in an axial--vector simplified model, as a function of
the mediator mass, for a model that assumes no tree-level
couplings to leptons. This plot is made for a particular choice of DM mass, 
but would look similar for other DM masses. 
Searches for boosted mediator decays are sensitive to masses as low as 50 GeV
and quark couplings \gq as low as 0.06 at 60 GeV. Jets from the
mediator decay are spatially separated for mediator masses above
250--300 GeV, where the $\gamma$ and gluon ISR+dijet channel
constrains \gq to be greater than 0.15--0.2. Above 400 GeV, where searches with
jets at the trigger level become available, they are the most
sensitive, excluding \gq as low as 0.05. Above 1 TeV, standard
dijet resonance and angular searches constrain quark couplings
from 0.1 to unity, up to 5 TeV.

Mixing between this mediator and the \textit{Z} boson induces loop-level
couplings to leptons. ATLAS and CMS use several sets of coupling
benchmarks to illustrate how the experimental constraints depend
on these unknown values. For equal couplings of the mediator to
leptons and jets, dilepton searches at a given mediator mass are
far more sensitive than dijet searches. Other values are
discussed further in the next section.

\subsubsection{Comparison of the sensitivity of visible and invisible LHC searches}\label{sub:comparisonVisibleInvisible}

Fully visible signatures of a particular dark interaction can be
powerful probes of it and, in some cases (e.g., when the {\IP}s
are too heavy to be directly produced), are the only way to observe
dark interactions at a collider. By contrast, only \MET
searches can observe invisible-particle production directly.
Each type of search complements the others; nevertheless, piecing
together searches in different channels requires a model.
Understanding precisely how these searches fit together can be
challenging when the model is uncertain.

As an example, we again consider the case of vector or
axial--vector mediators, to which both searches for anomalous \MET and two-body
resonance searches are sensitive. Although these models are simple,
they involved four parameters: two couplings, the invisible-particle
mass, and the mediator mass. Recent ATLAS and CMS results depict
results in a two-dimensional plane of mediator mass and DM
mass, following the recommendations of the LHC Dark Matter Working
Group.\footnote{Akin to simplified models of SUSY, where the axes
are neutralino mass and superpartner particle mass.}
Figure~\ref{fig:sensitivityComparison} shows a sample of
recent ATLAS plots. The remaining coupling parameters are
fixed to one of several benchmark sets that are selected on the basis of the
sensitivity of early Run 2 searches, precision constraints, and
the complementarity of different types of searches. Figure~\ref{fig:sensitivityComparison} displays the
constraints from dijet, dilepton, and \textit{X}+\MET searches on the
interaction model as excluded regions of the model
parameter space.

Figure~\ref{fig:sensitivityComparison}\textit{a} shows the LHC constraints
in a scenario that privileges dijet decays, for couplings
\gq = 0.25, \gl = 0, and \gdm = 1. 
In this case, dijet searches exclude
mediators between approximately 200~GeV and 2.6~TeV, while \textit{X}+\MET searches
can constrain even lighter mediators. 
Figure~\ref{fig:sensitivityComparison}\textit{b} shows the
exclusions for smaller quark couplings, \gq = 0.1, and a nonzero
lepton coupling, \gl = 0.01, chosen as indicative of the possible
size of loop-induced lepton couplings. With lower quark couplings,
and thus lower dijet production and decay rates, the regions of
masses excluded by the several dijet searches shrink. For
mediators heavier than 150 GeV, the exclusions from the recent
dilepton search fare better but do not extend very far into the
(smaller) region excluded by the jet+\MET search, where mediator
decays to DM dominate.

\begin{figure}[!htpb]
\includegraphics[width=0.75\textwidth]{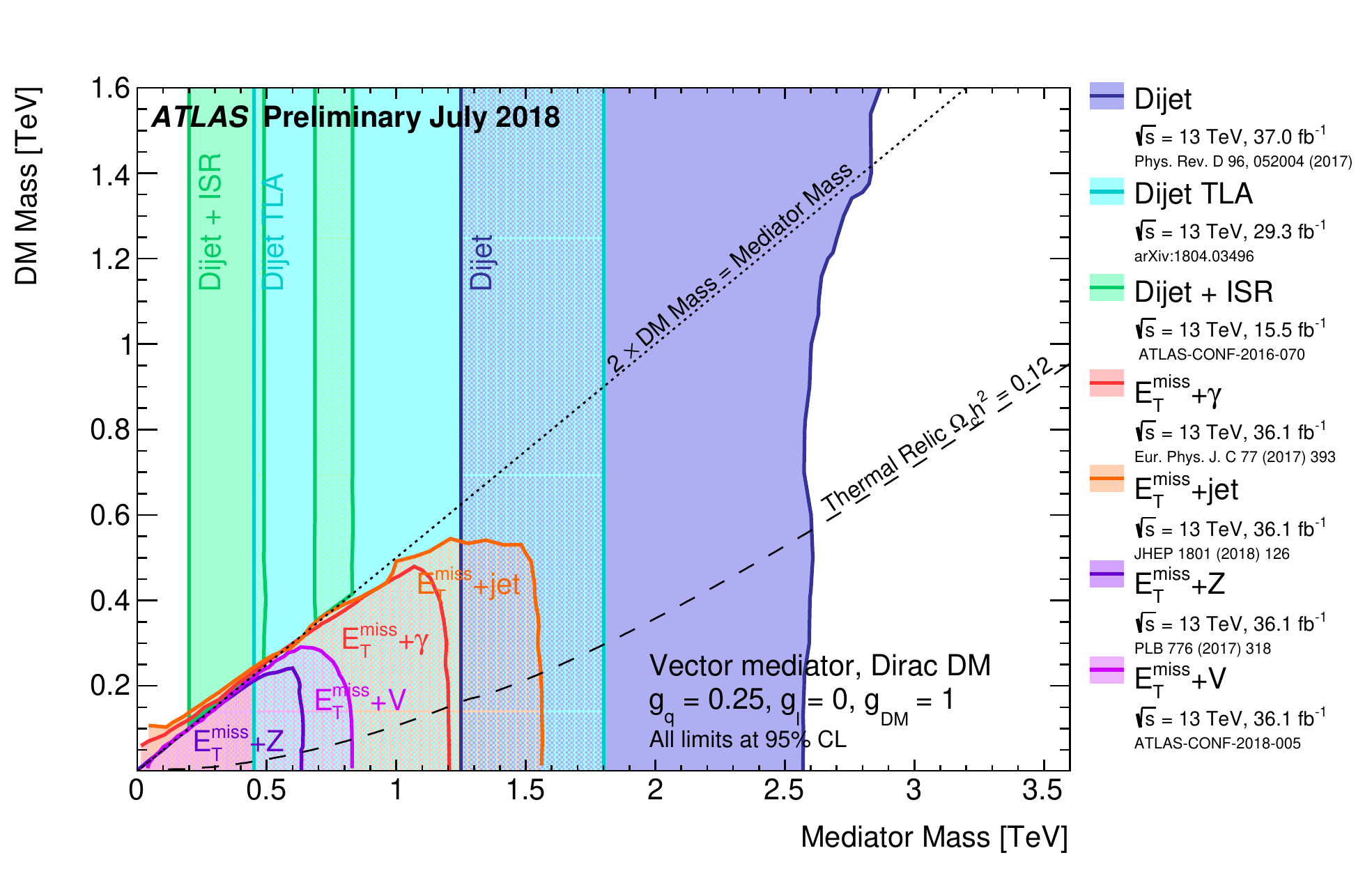}\\
\includegraphics[width=0.75\textwidth]{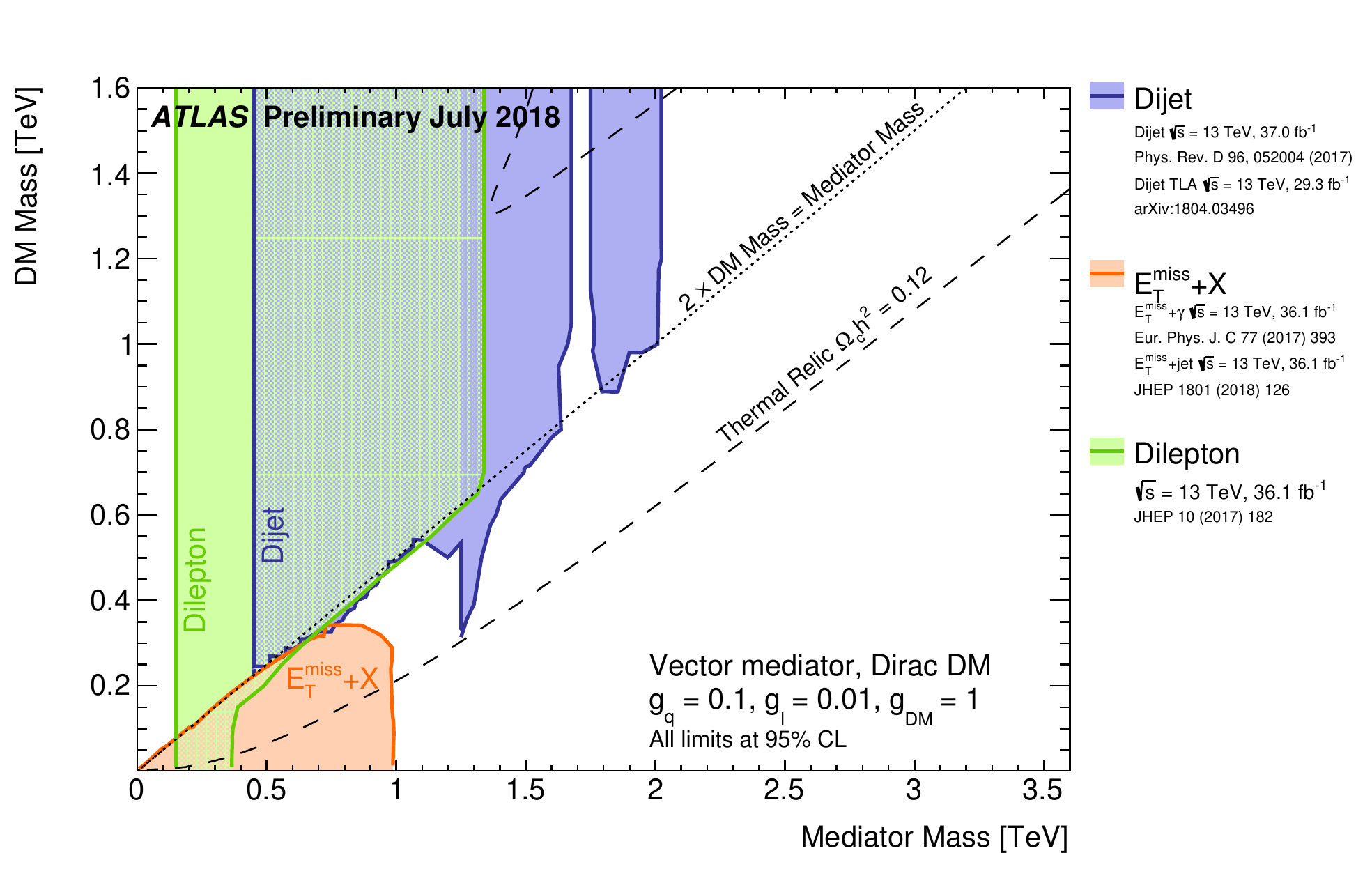}
\caption{Regions in DM mass--\Zprime mediator mass
excluded at 95\% CL by a selection of ATLAS searches 
(from References \cite{ATLAS:2016bvn,Aaboud:2018fzt,Aaboud:2017yvp,
Aaboud:2017phn,ATLAS-CONF-2018-005,Aaboud:2017bja,Aaboud:2017dor,
Aaboud:2017buh}) 
available as of July 2018, for two coupling scenarios. Dashed curves labeled ``thermal
relic'' indicate combinations of DM and mediator mass
that are consistent with a DM density of $\omega_c = 0.12
h^2$ and a standard thermal history, as computed in MadDM for this
model~\cite{Backovic:2015cra}. The dotted curve indicates the
kinematic threshold where the mediator can decay on-shell into
DM. 
In panel (a), the couplings of the mediator particle to each generation
of quarks (\gq) are set to 0.25, the couplings to leptons (\gl)
are set to zero and the coupling to DM is set to unity. 
In panel (b), \gq is set to 0.1, \gl is set to 0.01
and the coupling to DM \gdm is set to unity and marked as $g_{DM}$ in this plot.
Abbreviations:\ DM, dark matter; ISR, initial-state radiation; 
TLA, trigger-object level analysis. Adapted from~Reference \citen{ATLASSummary}.
}
\label{fig:sensitivityComparison}
\end{figure}

Thus, the relative sensitivity of visible and invisible searches
is both model and coupling dependent. One advantage of
searches for {\IP}s is their sensitivity to models
with very light mediators ($<$50 GeV) and on-shell decays to DM, since the
reach of dijet and dilepton searches to low-mass resonances is
still ultimately limited by data-taking constraints, 
but they are also stronger than direct mediator searches in the
case of supersymmetric models~\cite{Liew:2016oon}. 
Further examples of complementarity can be found in other types of models, 
for example, in SUSY, where striking signatures of visible particle decays (e.g. same-sign dileptons
or multiresonance final states) allow for discovery before more generic multijet + \MET signatures~\cite{Acharya:2009gb}. 

Since we do not know what DM model is realized in nature, 
all of these search channels are potentially relevant, 
and have different strengths. They are all necessary for 
a thorough search program. 

\section{COMPARISON OF COLLIDER RESULTS WITH DIRECT AND INDIRECT DETECTION EXPERIMENTS}\label{sec:04_Extrapolation}

A wide variety of reactions may produce {\IP}s at colliders, and
if the mediators of the interaction are light enough to be
produced on-shell, collider experiments are particularly well suited to
discovering and characterizing the interactions responsible.
Meanwhile, connecting a collider experiment's discovery or
nondiscovery of {\IP}s to DM requires both DD and
ID experiments, wherein Galactic DM
collides with a terrestrial target or extragalactic DM
annihilates.

Making this connection requires that one assume a particle
physics model. Within a given model and under well-specified
assumptions, the information obtained in a collider experiment can
be related to the information obtained in direct, indirect, and
astrophysical probes, and vice versa. One can then compare and
contrast the different types of information, for instance, to understand
where a DM discovery in current DD searches could be further
explored with mediator studies at the LHC, and where, among the
multitude of possible signals, collider searches might focus
their effort.

In the following subsections, we outline a strategy adopted by the ATLAS and
CMS experiments when making comparisons with astrophysical
observations (e.g., where a model is consistent with the present
DM density in the Universe) and with DD and ID results. We discuss the
assumptions made in the relic density calculation and in relating
reactions for {\IP}s to reactions of DM.

\subsection{Comparing LHC Constraints from Visible- and Invisible-Particle Searches with Noncollider Results}

The collider results from ATLAS and CMS typically appear as constraints
on production cross sections of specific processes, which are then
interpreted as statements about the fundamental parameters of a
simplified model (e.g., masses, couplings). Within the model,
information about the parameters can then be extrapolated to
statements about the noncollider observable of interest---for
example, the WIMP--nucleon scattering cross section for DD searches or the
thermal relic density. The LHC Dark
Matter Working Group (see Reference \citen{Boveia:2016mrp} and references therein) has provided instructions for how to
perform these extrapolations. For the first, partial Run 2 LHC search results, 
generic searches have only selected models that had an early chance of discovery 
for these comparisons. But many other
models can be used,\footnote{For reinterpretation of LHC results
and their comparisons to DD and ID searches for scalar and pseudoscalar
mediators, as well as in the context of 2HDMs, see, for example, References~\citen{Bell:2016ekl}, \citen{Athron:2017kgt}, and \citen{Banerjee:2017wxi}.} and
published searches typically provide some form of model-agnostic
results for this purpose.

For example, CMS and ATLAS have extrapolated the parameter exclusions
obtained by a recent set of searches to the spin-independent
WIMP--nucleon cross section of a DD experiment. The
result for CMS is depicted in Figure~\ref{fig:SICMS}, which shows a selection of DD results
for comparison. The figure illustrates general features of
many such comparisons. 

For spin-dependent DD scattering, such as an axial--vector mediated model, 
the LHC signals are relatively insensitive to the Lorentz structure of the 
interaction, while the DD signals are suppressed. 
As a result, the corresponding plots show that LHC searches play a more
powerful role relative to the DD searches over a wide range of invisible-particle masses. 
At intermediate DM masses,
both LHC and DD experiments have great potential for a discovery
and could verify one another's claims. 

One can also compare collider and ID results using
simplified model benchmarks. In traditional comparisons, only one
DM annihilation state at a time was used for the comparison
of collider and ID results, as in the case for $b\bar{b}$ (e.g., \citen{Agrawal:2014una}), but one can also compare ID and
LHC results for models annihilating to multiple final-state
fermions~\cite{Carpenter:2016thc}.

Recently, some DD and ID collaborations have adopted the benchmark
simplified models used by ATLAS and CMS (e.g.,
\citen{PhysRevLett.118.251301,Balazs:2017hxh}). IceCube and other
experiments have used constraints from a MSSM scan (e.g.,~\citen{Aartsen:2016zhm}). The pMSSM is also a good framework
to highlight the complementarity of LHC, DD, and ID experiments (e.g.,
\citen{Cahill-Rowley:2014twa}).

\begin{marginnote}[]
\entry{The LHC Dark Matter Working Group}{provides for the
translation of LHC limits to DD and ID searches~\cite{Boveia:2016mrp},
as well as calculations of relic density
(see \url{https://gitlab.cern.ch/lhc-dmwg-material/relic-density})}
\end{marginnote}

We emphasize that the exclusion regions obtained in this
way will depend strongly on the assumptions of the model. The
extrapolations are done with full knowledge that the simplified
model is merely a crude guess, and one must be careful not to
overgeneralize. Neither this procedure nor the
simplified models themselves account for effects outside the
model, such as interference and mixing with Standard Model boson and quarkonia
resonances, or the evolution of the operators in the model from
the LHC collision energies to other energy
scales~\cite{DEramo:2014nmf}. Moreover, all experimental results,
be they from DD, ID, or collider searches, are affected by experimental and
theoretical uncertainties not shown in Figure~\ref{fig:SICMS}.
In principle, LHC \MET searches cannot probe cases where 
WIMPs are so strongly interacting that they are
stopped in the detector, see e.g.~\cite{Daci:2015hca,Cappiello:2018hsu}. 
LHC results haven't yet explicitly quoted an upper range to
their bounds in Figure~\ref{fig:SICMS}. 

\begin{figure}[!htpb]
\includegraphics[width=0.9\textwidth]{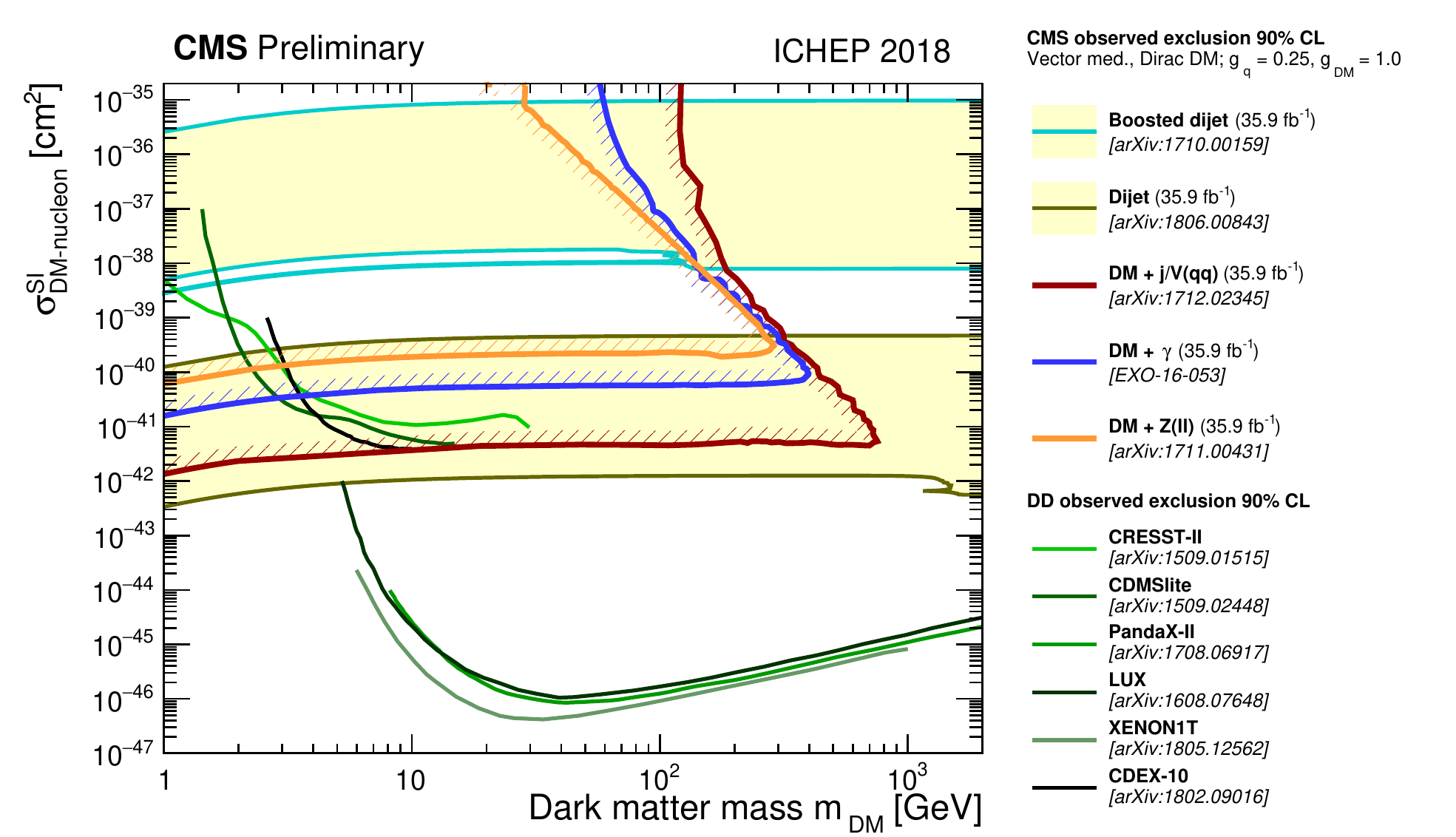}
\caption{The 90\%-CL constraints from the CMS experiment 
from References \cite{Sirunyan:2017nvi,Sirunyan:2018xlo,Sirunyan:2017jix,CMS-PAS-EXO-16-053,Sirunyan:2017qfc}
in the \mdm-spin-independent DM--nucleon plane for a vector mediator,
Dirac DM, and benchmark couplings \gq = 0.25 and \gdm = 1.0 (marked as $g_{DM}$ in this plot) chosen as an example of what
early LHC searches would be sensitive to, compared with direct detection
experiments from References \cite{Angloher:2015ewa,Agnese:2015nto,Cui:2017nnn,Akerib:2016vxi,Aprile:2018dbl}. 
It is important to note that this comparison is only valid for this particular
combination of model and parameter choices. 
Abbreviation:\ DM, dark matter. Adapted from Reference~\citen{CMSSummary}.} \label{fig:SICMS}
\end{figure}

\subsection{Relic Density Considerations}

In the absence of a signal in noncollider experiments, the ability of a
model to link its {\IP}s with the observed DM abundance is key to
distinguishing it from other types of models of BSM\ physics. Making this link, however, requires extrapolating
from the present day to the early Universe along an increasingly
tenuous chain of assumptions. For simplified models, this is
especially problematic, because the model is designed to describe
collider-scale processes; it may not even contain higher-scale interactions
relevant in the early Universe.

Nevertheless, it is interesting to
examine the parameter regions in models that can make the link, even if in limited
situations~\cite{Busoni:2014gta,Catena:2017xqq}. For example, for
the general simplified models discussed in
Section~\ref{sub:simplifiedModels}, one can use programs
such as MadDM and
MicrOMEGas~\cite{Backovic:2015cra,Barducci:2016pcb} to compute  the DM
abundance for a standard thermal relic, assuming that the
interaction described by the simplified model is the one
responsible for setting the relic density. Often (e.g.,
Figure~\ref{fig:sensitivityComparison}), ATLAS and CMS supplement
their results with contours indicating where within a model this
procedure obtains the correct DM density of $\omega_c =
0.12 h^2$. These lines should be regarded as guidance rather than as strict
requirements for the models considered. 
When the model cannot reproduce the correct abundance indicates either that  the model requires additional
components beyond those included in the simplified model or that
the chain of assumptions is incorrect~\cite{Bernal:2017kxu}.

\section{OUTLOOK}\label{sec:05_Future}

We are optimistic that, in the next decade, the variety of powerful searches for
particle DM, from collider approaches to underground
experiments to observatories, will lead to much progress toward its eventual discovery.
Collider strategies that have been essential to discover and understand the fundamental particles
of the Standard Model are being extended with new techniques designed to extract rare and difficult
signals from the data. Our understanding of
collider search targets is also rapidly improving. Along with SUSY
benchmarks motivated by the hierarchy problem, targets directly
motivated by DM observations are encouraging a new generation of
experimentalists to branch out into directions that so far have been only
sparsely explored. Finally, the HL-LHC data set will exceed that presented here by a factor of 100. These
are exciting times~\cite{Steigman:1979kw}.

\begin{issues}[FUTURE ISSUES]
\begin{enumerate}

\item The obvious directions for LHC searches are toward lower-rate
processes and processes that are more difficult to detect:

\begin{itemize}

\item Extended scalar sectors and electroweak SUSY
are among the possible benchmarks for lower rate processes.
As the LHC experiments record progressively more data during Run 2 and Run 3, 
searches are becoming sensitive to the simplest scalar simplified
models, opening the door to more realistic models. On a longer
timescale, the HL-LHC data set will bring sensitivity up to 3
TeV in scalar mediator masses for unit
couplings~\cite{CMS-PAS-FTR-16-005} and precision knowledge of the
Standard Model Higgs sector, and the mass reach for
electroweak production of SUSY partners should increase by a few hundred
GeV~\cite{Campana:2016cqm}.

\item With data arriving at a slightly less frantic pace, experimentally
challenging LLP signatures are a growing field,
and benchmarks similar to Reference~\citen{Abercrombie:2015wmb} are
needed to help guide the design of these searches. Among many
ongoing efforts are the bottom-up approach adopted in
Reference \citen{Buchmueller:2017uqu}, which connects such models with
the LLP limit of those described in
Section~\ref{sub:simplifiedModels}. 
Many LLP searches have not yet been done, and not all existing searches have been optimized. 
Therefore, LLP searches have the potential for substantial improvements, 
much beyond those expected by the accumulation of luminosity. 

\end{itemize}

\item Precision searches, detector upgrades, and efficient triggering of rare signals buried
in large backgrounds are key to fully exploiting the HL-LHC data set: 

\begin{itemize} 

\item In the familiar jet+\MET search,
precision estimates of the \textit{V}+jet backgrounds, and of the inputs to
these predictions, will be crucial. Efforts in that
direction are ongoing \cite{Blumenschein:2018gtm}.

\item Upgrades for Run 3 and HL-LHC provide new capabilities that may make
new data more valuable for these searches than what recorded so far
for rare processes involving light new particles. This is a subject that has been largely
unexplored for ATLAS and CMS~\cite{Alves:2016cqf} and that can be developed
further when tracking information is available at the trigger level to remove pileup.
LHCb will make use of a novel triggerless detector readout to
perform dark photon searches with unprecedented
sensitivities~\cite{Ilten:2016tkc}. 

\end{itemize}

\item Future hadron and electron--positron colliders have immense
potential [see, e.g., studies on a future hadron
collider~\cite{Golling:2016gvc}]. Nevertheless, present studies
largely continue the approaches already in use at the LHC. A new
hadron collider would be built to discover New Physics;
therefore, qualitatively different experimental design, benchmark
models, and analysis strategies should be considered.

\item More useful working comparisons between results from colliders,
underground searches, and observatories should take into account
the uncertainties on each type of result, and on the
extrapolations between them. The main uncertainties for LHC
searches are outlined in
Section~\ref{sec:03_ExperimentalResults} and in the experimental
references; for a summary of DD and ID uncertainties, see
References~\citen{Feldstein:2014ufa} and \citen{d300ef23986a49099715e661295a4d72} and
references therein.

\item Comparisons among collider and noncollider particle physics
experiments are becoming standard; relating particle physics to
astrophysical observables is crucial to exploit the few clues that
DM can provide about BSM\ particle physics. We
strongly encourage further research on this subject (e.g.,
\citen{Buckley:2017ijx}).
\end{enumerate}
\end{issues}

\section*{DISCLOSURE STATEMENT}

The authors are not aware of any affiliations, memberships,
funding, or financial holdings that might be perceived as
affecting the objectivity of this review.

\section*{ACKNOWLEDGMENTS}

We thank William Kalderon for the help in preparation of Fig. 6. We also thank Teng Jian Khoo, Suchita Kulkarni, Robert Harris, Priscilla Pani, Maurizio Pierini, Frederik Ruehr, Tim Tait, Emma Tolley, Liantao Wang, Tyler Wang, and Bryan Zaldivar for their help and advice in preparing this manuscript. We also thank Ulrich Haisch, Valerio Ippolito, Christian Ohm, Jessie Shelton, and Mike Williams for useful discussion. Research by A.B. is supported by the US Department of Energy (grant DE-SC0011726). Research by C.D. is part of a project that has received funding from the European Research Council under the European Union's Horizon 2020 research and innovation program (grant agreement 679305) and from the Swedish Research Council.

\section*{RELATED RESOURCES}
\begin{enumerate}
\item Bertone G, et al. \textit{Particle Dark Matter: Observations, Models and Searches}. Cambridge, UK: Cambridge Univ. Press (2010)%}

\item Bertone G, Hooper D. \href{https://arxiv.org/abs/1605.04909}{arXiv:1605.04909} [astro-ph.CO] (2016)

\item Plehn T. \textit{Yet another introduction to dark matter}. Lect. notes, Inst. Theor. Phys., Univ. Heidelberg, Ger. \url{http://www.thphys.uni-heidelberg.de/~plehn/pics/dark_matter.pdf} (2017)

\item Arcadi G, et al. \textit{Eur. Phys. J.} \textit{C} 78:203 (2018) \href{https://arxiv.org/abs/1703.07364}{arXiv:1703.07364}

\item Kahlhoefer F.  \textit{Int. J. Mod. Phys. A} 32:1730006 (2017) \href{https://arxiv.org/abs/1702.02430}{arXiv:1702.02430}

\item {LHC Physics Center CERN.\textit{ WG on dark matter searches at the LHC}. DMWG Work. Group home page, CERN, Geneva.} \url{http://lpcc.web.cern.ch/content/lhc-dm-wg-wg-dark-matter-searches-lhc} (2015)

\end{enumerate}
\end{document}